%%%%%%%%%%%%%%%%%%  tex macros for preprints, cm version %%%%%%%%%%%%%%
%         (P. Ginsparg <ginsparg@lanl.gov>, last updated 7/94)
%                if confused, type `b' in response to query 
%           hypertex extensions (still provisional), 7/26/94
%
%---------------------------------------------------------------------%
\input hyperbasics %comment out this line to restore non-hyper functionality
%
%% site dependent options:
%% \unredoffs and \redoffs define horizontal and vertical offsets
%% respectively for unreduced and reduced modes. \speclscape defines
%% the \special{} call that sets printer to landscape (sideways) mode.
%% from standard set below, leave uncommented as appropriate or redefine
%
%%% next 400dpi
\def\unredoffs{} \def\redoffs{\voffset=-.31truein\hoffset=-.48truein}
\def\speclscape{}
%\def\speclscape{\special{papersize=11in,8.5in}}
%
%%% apple lw
%\def\unredoffs{} \def\redoffs{\voffset=-.31truein\hoffset=-.59truein}
%\def\speclscape{\special{ps: landscape}}
%
%%% qms lasergrafix:
%\def\unredoffs{} \def\redoffs{\voffset=-.4truein\hoffset=.125truein}
%\def\speclscape{\special{qms: landscape}}
%
%%% saclay A4 paper:
%\def\unredoffs{\hoffset-.14truein\voffset-.2truein}
%\def\redoffs{\voffset=-.45truein\hoffset=-.21truein}
%\def\speclscape{\special{landscape}}
%
%---------------------------------------------------------------------%
%
\newbox\leftpage \newdimen\fullhsize \newdimen\hstitle \newdimen\hsbody
\tolerance=1000\hfuzz=2pt
\catcode`\@=11 % This allows us to modify PLAIN macros.
\ifx\hyperdef\UNd@FiNeD\def\hyperdef#1#2#3#4{#4}\def\hyperref#1#2#3#4{#4}\fi
\def\bigans{b }
\def\answ{b }
%\message{ big or little (b/l)? }\read-1 to\answ
%
\ifx\answ\bigans\message{(This will come out unreduced.}
\magnification=1200\unredoffs\baselineskip=16pt plus 2pt minus 1pt
\hsbody=\hsize \hstitle=\hsize %take default values for unreduced format
\else\message{(This will be reduced.} \let\l@r=L
\magnification=1000\baselineskip=16pt plus 2pt minus 1pt \vsize=7truein
\redoffs \hstitle=8truein\hsbody=4.75truein\fullhsize=10truein\hsize=\hsbody
\output={\ifnum\pageno=0 %%% This is the HUTP version
  \shipout\vbox{\speclscape{\hsize\fullhsize\makeheadline}
    \hbox to \fullhsize{\hfill\pagebody\hfill}}\advancepageno
  \else
  \almostshipout{\leftline{\vbox{\pagebody\makefootline}}}\advancepageno
  \fi}
\def\almostshipout#1{\if L\l@r \count1=1 \message{[\the\count0.\the\count1]}
      \global\setbox\leftpage=#1 \global\let\l@r=R
 \else \count1=2
  \shipout\vbox{\speclscape{\hsize\fullhsize\makeheadline}
      \hbox to\fullhsize{\box\leftpage\hfil#1}}  \global\let\l@r=L\fi}
\fi
%---------------------------------------------------------------------
%
\newcount\yearltd\yearltd=\year%\advance\yearltd by -1900

\def\Title#1#2{\nopagenumbers\abstractfont\hsize=\hstitle\rightline{#1}%
\vskip 1in\centerline{\titlefont #2}\abstractfont\vskip .5in\pageno=0}
\def\Date#1{\vfill\leftline{#1}\tenpoint\supereject\global\hsize=\hsbody%
\footline={\hss\tenrm\hyperdef\hypernoname{page}\folio\folio\hss}}%
% (restores pagenumbers)
%
%       use following instead of \Date on the preliminary draft,
%       puts date/time on each page in big mode, writes labels in margins

\def\draftmode{\message{ DRAFTMODE }\def\draftdate{{\rm preliminary draft:
\number\month/\number\day/\number\yearltd\ \ \hourmin}}%
\headline={\hfil\draftdate}\writelabels\baselineskip=20pt plus 2pt minus 2pt
 {\count255=\time\divide\count255 by 60 \xdef\hourmin{\number\count255}
  \multiply\count255 by-60\advance\count255 by\time
  \xdef\hourmin{\hourmin:\ifnum\count255<10 0\fi\the\count255}}}
%       use \nolabels to get rid of eqn, ref, and fig labels in draft mode
\def\nolabels{\def\wrlabeL##1{}\def\eqlabeL##1{}\def\reflabeL##1{}}
\def\writelabels{\def\wrlabeL##1{\leavevmode\vadjust{\rlap{\smash%
{\line{{\escapechar=` \hfill\rlap{\sevenrm\hskip.03in\string##1}}}}}}}%
\def\eqlabeL##1{{\escapechar-1\rlap{\sevenrm\hskip.05in\string##1}}}%
\def\reflabeL##1{\noexpand\llap{\noexpand\sevenrm\string\string\string##1}}}
\nolabels
%
% tagged sec numbers
\global\newcount\secno \global\secno=0
\global\newcount\meqno \global\meqno=1
\def\s@csym{}
\def\newsec#1{\global\advance\secno by1%
{\toks0{#1}\message{(\the\secno. \the\toks0)}}%
%\ifx\answ\bigans \vfill\eject \else \bigbreak\bigskip \fi  %if desired
\global\subsecno=0\eqnres@t\let\s@csym\secsym\xdef\secn@m{\the\secno}\noindent
{\bf\hyperdef\hypernoname{section}{\the\secno}{\the\secno.} #1}%
\writetoca{{\string\hyperref{}{section}{\the\secno}{\the\secno.}} {\bf {#1}}}%
\par\nobreak\medskip\nobreak}
\def\eqnres@t{\xdef\secsym{\the\secno.}\global\meqno=1\bigbreak\bigskip}
\def\sequentialequations{\def\eqnres@t{\bigbreak}}\xdef\secsym{}
\global\newcount\subsecno \global\subsecno=0
\def\subsec#1{\global\advance\subsecno by1%
{\toks0{#1}\message{(\s@csym\the\subsecno. \the\toks0)}}%
\ifnum\lastpenalty>9000\else\bigbreak\fi
\noindent{\it\hyperdef\hypernoname{subsection}{\secn@m.\the\subsecno}%
{\secn@m.\the\subsecno.} #1}\writetoca{\string\quad
{\string\hyperref{}{subsection}{\secn@m.\the\subsecno}{\secn@m.\the\subsecno.}}
{#1}}\par\nobreak\medskip\nobreak}
\def\appendix#1#2{\global\meqno=1\global\subsecno=0\xdef\secsym{\hbox{#1.}}%
\bigbreak\bigskip\noindent{\bf Appendix \hyperdef\hypernoname{appendix}{#1}%
{#1.} #2}{\toks0{(#1. #2)}\message{\the\toks0}}%
\xdef\s@csym{#1.}\xdef\secn@m{#1}%
\writetoca{\string\hyperref{}{appendix}{#1}{Appendix {#1.}} {#2}}%
\par\nobreak\medskip\nobreak}
%
%       \eqn\label{a+b=c}	gives displayed equation, numbered
%				consecutively within sections.
%     \eqnn and \eqna define labels in advance (of eqalign?)
%
\def\checkm@de#1#2{\ifmmode{\def\f@rst##1{##1}\hyperdef\hypernoname{equation}%
{#1}{#2}}\else\hyperref{}{equation}{#1}{#2}\fi}
\def\eqnn#1{\DefWarn#1\xdef #1{(\noexpand\relax\noexpand\checkm@de%
{\s@csym\the\meqno}{\secsym\the\meqno})}%
\wrlabeL#1\writedef{#1\leftbracket#1}\global\advance\meqno by1}
\def\f@rst#1{\c@t#1a\em@ark}\def\c@t#1#2\em@ark{#1}
\def\eqna#1{\DefWarn#1\wrlabeL{#1$\{\}$}%
\xdef #1##1{(\noexpand\relax\noexpand\checkm@de%
{\s@csym\the\meqno\noexpand\f@rst{##1}}{\hbox{$\secsym\the\meqno##1$}})}
\writedef{#1\numbersign1\leftbracket#1{\numbersign1}}\global\advance\meqno by1}
\def\eqn#1#2{\DefWarn#1%
\xdef #1{(\noexpand\hyperref{}{equation}{\s@csym\the\meqno}%
{\secsym\the\meqno})}$$#2\eqno(\hyperdef\hypernoname{equation}%
{\s@csym\the\meqno}{\secsym\the\meqno})\eqlabeL#1$$%
\writedef{#1\leftbracket#1}\global\advance\meqno by1}
\def\xeqn{\expandafter\xe@n}\def\xe@n(#1){#1}
\def\xeqna#1{\expandafter\xe@n#1}
\def\eqns#1{(\e@ns #1{\hbox{}})}
\def\e@ns#1{\ifx\UNd@FiNeD#1\message{eqnlabel \string#1 is undefined.}%
\xdef#1{(?.?)}\fi{\let\hyperref=\relax\xdef\next{#1}}%
\ifx\next\em@rk\def\next{}\else%
\ifx\next#1\xeqn#1\else\def\n@xt{#1}\ifx\n@xt\next#1\else\xeqna#1\fi
\fi\let\next=\e@ns\fi\next}

\def\DefWarn#1{\ifx\UNd@FiNeD#1\else
\immediate\write16{*** WARNING: the label \string#1 is already defined ***}\fi}
%
%			 footnotes
\newskip\footskip\footskip14pt plus 1pt minus 1pt %sets footnote baselineskip
\def\footnotefont{\ninepoint}\def\f@t#1{\footnotefont #1\@foot}
\def\f@@t{\baselineskip\footskip\bgroup\footnotefont\aftergroup\@foot\let\next}
\setbox\strutbox=\hbox{\vrule height9.5pt depth4.5pt width0pt}
\global\newcount\ftno \global\ftno=0
\def\foot{\global\advance\ftno by1\def\foot@rg{\hyperref{}{footnote}%
{\the\ftno}{\the\ftno}\xdef\foot@rg{\noexpand\hyperdef\noexpand\hypernoname%
{footnote}{\the\ftno}{\the\ftno}}}\footnote{$^{\foot@rg}$}}
%
%say \footend to put footnotes at end
%will cause problems if \ref used inside \foot, instead use \nref before
\newwrite\ftfile
\def\footend{\def\foot{\global\advance\ftno by1\chardef\wfile=\ftfile
%%$^{\the\ftno}$\ifnum\ftno=1\immediate\openout\ftfile=\jobname.fts\fi%
\hyperref{}{footnote}{\the\ftno}{$^{\the\ftno}$}%
\ifnum\ftno=1\immediate\openout\ftfile=\jobname.fts\fi%
\immediate\write\ftfile{\noexpand\smallskip%
%%\noexpand\item{f\the\ftno:\ }\pctsign}\findarg}%
\noexpand\item{\noexpand\hyperdef\noexpand\hypernoname{footnote}
{\the\ftno}{f\the\ftno}:\ }\pctsign}\findarg}%
\def\footatend{\vfill\eject\immediate\closeout\ftfile{\parindent=20pt
\centerline{\bf Footnotes}\nobreak\bigskip\input \jobname.fts }}}
\def\footatend{}
%
%     \ref\label{text}
% generates a number, assigns it to \label, generates an entry.
% To list the refs on a separate page,  \listrefs
%
\global\newcount\refno \global\refno=1
\newwrite\rfile
\def\ref{[\hyperref{}{reference}{\the\refno}{\the\refno}]\nref}
\def\nref#1{\DefWarn#1%
\xdef#1{[\noexpand\hyperref{}{reference}{\the\refno}{\the\refno}]}%
\writedef{#1\leftbracket#1}%
\ifnum\refno=1\immediate\openout\rfile=\jobname.refs\fi
\chardef\wfile=\rfile\immediate\write\rfile{\noexpand\item{[\noexpand\hyperdef%
\noexpand\hypernoname{reference}{\the\refno}{\the\refno}]\ }%
\reflabeL{#1\hskip.31in}\pctsign}\global\advance\refno by1\findarg}
%	horrible hack to sidestep tex \write limitation
\def\findarg#1#{\begingroup\obeylines\newlinechar=`\^^M\pass@rg}
{\obeylines\gdef\pass@rg#1{\writ@line\relax #1^^M\hbox{}^^M}%
\gdef\writ@line#1^^M{\expandafter\toks0\expandafter{\striprel@x #1}%
\edef\next{\the\toks0}\ifx\next\em@rk\let\next=\endgroup\else\ifx\next\empty%
\else\immediate\write\wfile{\the\toks0}\fi\let\next=\writ@line\fi\next\relax}}
\def\striprel@x#1{} \def\em@rk{\hbox{}}
\def\lref{\begingroup\obeylines\lr@f}
\def\lr@f#1#2{\DefWarn#1\gdef#1{\let#1=\UNd@FiNeD\ref#1{#2}}\endgroup\unskip}

\def\addref#1{\immediate\write\rfile{\noexpand\item{}#1}} %now unnecessary
\def\listrefs{\footatend\immediate\closeout\rfile\writestoppt
\baselineskip=\footskip\centerline{{\bf References}}\bigskip{\parindent=20pt%
\frenchspacing\escapechar=` \input \jobname.refs\vfill\eject}\nonfrenchspacing}
\def\startrefs#1{\immediate\openout\rfile=\jobname.refs\refno=#1}
\def\xref{\expandafter\xr@f}\def\xr@f[#1]{#1}
\def\refs#1{\count255=1[\r@fs #1{\hbox{}}]}
\def\r@fs#1{\ifx\UNd@FiNeD#1\message{reflabel \string#1 is undefined.}%
\nref#1{need to supply reference \string#1.}\fi%
\vphantom{\hphantom{#1}}{\let\hyperref=\relax\xdef\next{#1}}%
\ifx\next\em@rk\def\next{}%
\else\ifx\next#1\ifodd\count255\relax\xref#1\count255=0\fi%
\else#1\count255=1\fi\let\next=\r@fs\fi\next}
%

%
% this is ugly, but moore insists
\newwrite\ffile\global\newcount\figno \global\figno=1
\def\fig{fig.~\hyperref{}{figure}{\the\figno}{\the\figno}\nfig}
\def\nfig#1{\DefWarn#1%
\xdef#1{fig.~\noexpand\hyperref{}{figure}{\the\figno}{\the\figno}}%
\writedef{#1\leftbracket fig.\noexpand~\xfig#1}%
\ifnum\figno=1\immediate\openout\ffile=\jobname.figs\fi\chardef\wfile=\ffile%
{\let\hyperref=\relax
\immediate\write\ffile{\noexpand\medskip\noexpand\item{Fig.\ %
\noexpand\hyperdef\noexpand\hypernoname{figure}{\the\figno}{\the\figno}. }
\reflabeL{#1\hskip.55in}\pctsign}}\global\advance\figno by1\findarg}
\def\listfigs{\vfill\eject\immediate\closeout\ffile{\parindent40pt
\baselineskip14pt\centerline{{\bf Figure Captions}}\nobreak\medskip
\escapechar=` \input \jobname.figs\vfill\eject}}
\def\xfig{\expandafter\xf@g}\def\xf@g fig.\penalty\@M\ {}
\def\figs#1{figs.~\f@gs #1{\hbox{}}}
\def\f@gs#1{{\let\hyperref=\relax\xdef\next{#1}}\ifx\next\em@rk\def\next{}\else
\ifx\next#1\xfig #1\else#1\fi\let\next=\f@gs\fi\next}
\def\figin{\epsfcheck\figin}\def\figins{\epsfcheck\figins}
\def\epsfcheck{\ifx\epsfbox\UNd@FiNeD
\message{(NO epsf.tex, FIGURES WILL BE IGNORED)}
\gdef\figin##1{\vskip2in}\gdef\figins##1{\hskip.5in}% blank space instead
\else\message{(FIGURES WILL BE INCLUDED)}%
\gdef\figin##1{##1}\gdef\figins##1{##1}\fi}
\def\DefWarn#1{}
\def\figinsert{\goodbreak\midinsert}
\def\ifig#1#2#3{\DefWarn#1\xdef#1{fig.~\noexpand\hyperref{}{figure}%
{\the\figno}{\the\figno}}\writedef{#1\leftbracket fig.\noexpand~\xfig#1}%
\figinsert\figin{\centerline{#3}}\medskip\centerline{\vbox{\baselineskip12pt
\advance\hsize by -1truein\noindent\wrlabeL{#1=#1}\footnotefont%
{\bf Fig.~\hyperdef\hypernoname{figure}{\the\figno}{\the\figno}:} #2}}
\bigskip\endinsert\global\advance\figno by1}
\newwrite\lfile
{\escapechar-1\xdef\pctsign{\string\%}\xdef\leftbracket{\string\{}
\xdef\rightbracket{\string\}}\xdef\numbersign{\string\#}}
\def\writedefs{\immediate\openout\lfile=\jobname.defs \def\writedef##1{%
{\let\hyperref=\relax\let\hyperdef=\relax\let\hypernoname=\relax
 \immediate\write\lfile{\string\def\string##1\rightbracket}}}}%
\def\writestop{\def\writestoppt{\immediate\write\lfile{\string\pageno
 \the\pageno\string\startrefs\leftbracket\the\refno\rightbracket
 \string\def\string\secsym\leftbracket\secsym\rightbracket
 \string\secno\the\secno\string\meqno\the\meqno}\immediate\closeout\lfile}}
\def\writestoppt{}\def\writedef#1{}
\def\seclab#1{\DefWarn#1%
\xdef #1{\noexpand\hyperref{}{section}{\the\secno}{\the\secno}}%
\writedef{#1\leftbracket#1}\wrlabeL{#1=#1}}
\def\subseclab#1{\DefWarn#1%
\xdef #1{\noexpand\hyperref{}{subsection}{\secn@m.\the\subsecno}%
{\secn@m.\the\subsecno}}\writedef{#1\leftbracket#1}\wrlabeL{#1=#1}}
\def\applab#1{\DefWarn#1%
\xdef #1{\noexpand\hyperref{}{appendix}{\secn@m}{\secn@m}}%
\writedef{#1\leftbracket#1}\wrlabeL{#1=#1}}
\newwrite\tfile \def\writetoca#1{}
\def\leaderfill{\leaders\hbox to 1em{\hss.\hss}\hfill}
%	use this to write file with table of contents
\def\writetoc{\immediate\openout\tfile=\jobname.toc
   \def\writetoca##1{{\edef\next{\write\tfile{\noindent ##1
   \string\leaderfill {\string\hyperref{}{page}{\noexpand\number\pageno}%
                       {\noexpand\number\pageno}} \par}}\next}}}
%       and this lists table of contents on second pass
\newread\ch@ckfile
\def\listtoc{\immediate\closeout\tfile\immediate\openin\ch@ckfile=\jobname.toc
\ifeof\ch@ckfile\message{no file \jobname.toc, no table of contents this pass}%
\else\closein\ch@ckfile\centerline{\bf Contents}\nobreak\medskip%
{\baselineskip=12pt\footnotefont\parskip=0pt\catcode`\@=11\input\jobname.toc
\catcode`\@=12\bigbreak\bigskip}\fi}
\catcode`\@=12 % at signs are no longer letters
%
%	Unpleasantness in calling in abstract and title fonts
\edef\tfontsize{\ifx\answ\bigans scaled\magstep3\else scaled\magstep4\fi}
\font\titlerm=cmr10 \tfontsize \font\titlerms=cmr7 \tfontsize
\font\titlermss=cmr5 \tfontsize \font\titlei=cmmi10 \tfontsize
\font\titleis=cmmi7 \tfontsize \font\titleiss=cmmi5 \tfontsize
\font\titlesy=cmsy10 \tfontsize \font\titlesys=cmsy7 \tfontsize
\font\titlesyss=cmsy5 \tfontsize \font\titleit=cmti10 \tfontsize
\skewchar\titlei='177 \skewchar\titleis='177 \skewchar\titleiss='177
\skewchar\titlesy='60 \skewchar\titlesys='60 \skewchar\titlesyss='60
\def\titlefont{\def\rm{\fam0\titlerm}% switch to title font
\textfont0=\titlerm \scriptfont0=\titlerms \scriptscriptfont0=\titlermss
\textfont1=\titlei \scriptfont1=\titleis \scriptscriptfont1=\titleiss
\textfont2=\titlesy \scriptfont2=\titlesys \scriptscriptfont2=\titlesyss
\textfont\itfam=\titleit \def\it{\fam\itfam\titleit}\rm}
 \ifx\answ\bigans\else scaled\magstep1\fi
\ifx\answ\bigans\def\abstractfont{\tenpoint}\else
\font\absit=cmti10 scaled \magstep1
\font\abssl=cmsl10 scaled \magstep1
\font\absrm=cmr10 scaled\magstep1 \font\absrms=cmr7 scaled\magstep1
\font\absrmss=cmr5 scaled\magstep1 \font\absi=cmmi10 scaled\magstep1
\font\absis=cmmi7 scaled\magstep1 \font\absiss=cmmi5 scaled\magstep1
\font\abssy=cmsy10 scaled\magstep1 \font\abssys=cmsy7 scaled\magstep1
\font\abssyss=cmsy5 scaled\magstep1 \font\absbf=cmbx10 scaled\magstep1
\skewchar\absi='177 \skewchar\absis='177 \skewchar\absiss='177
\skewchar\abssy='60 \skewchar\abssys='60 \skewchar\abssyss='60
\def\abstractfont{\def\rm{\fam0\absrm}% switch to abstract font
\textfont0=\absrm \scriptfont0=\absrms \scriptscriptfont0=\absrmss
\textfont1=\absi \scriptfont1=\absis \scriptscriptfont1=\absiss
\textfont2=\abssy \scriptfont2=\abssys \scriptscriptfont2=\abssyss
\textfont\itfam=\absit \def\it{\fam\itfam\absit}\def\footnotefont{\tenpoint}%
\textfont\slfam=\abssl \def\sl{\fam\slfam\abssl}%
\textfont\bffam=\absbf \def\bf{\fam\bffam\absbf}\rm}\fi
\def\tenpoint{\def\rm{\fam0\tenrm}% switch back to 10-point type
\textfont0=\tenrm \scriptfont0=\sevenrm \scriptscriptfont0=\fiverm
\textfont1=\teni  \scriptfont1=\seveni  \scriptscriptfont1=\fivei
\textfont2=\tensy \scriptfont2=\sevensy \scriptscriptfont2=\fivesy
\textfont\itfam=\tenit \def\it{\fam\itfam\tenit}\def\footnotefont{\ninepoint}%
\textfont\bffam=\tenbf \def\bf{\fam\bffam\tenbf}\def\sl{\fam\slfam\tensl}\rm}
\font\ninerm=cmr9 \font\sixrm=cmr6 \font\ninei=cmmi9 \font\sixi=cmmi6
\font\ninesy=cmsy9 \font\sixsy=cmsy6 \font\ninebf=cmbx9
\font\nineit=cmti9 \font\ninesl=cmsl9 \skewchar\ninei='177
\skewchar\sixi='177 \skewchar\ninesy='60 \skewchar\sixsy='60
\def\ninepoint{\def\rm{\fam0\ninerm}% switch to footnote font
\textfont0=\ninerm \scriptfont0=\sixrm \scriptscriptfont0=\fiverm
\textfont1=\ninei \scriptfont1=\sixi \scriptscriptfont1=\fivei
\textfont2=\ninesy \scriptfont2=\sixsy \scriptscriptfont2=\fivesy
\textfont\itfam=\ninei \def\it{\fam\itfam\nineit}\def\sl{\fam\slfam\ninesl}%
\textfont\bffam=\ninebf \def\bf{\fam\bffam\ninebf}\rm}
%
%---------------------------------------------------------------------
%
\def\noblackbox{\overfullrule=0pt}
\hyphenation{anom-aly anom-alies coun-ter-term coun-ter-terms}
\def\inv{^{\raise.15ex\hbox{${\scriptscriptstyle -}$}\kern-.05em 1}}

\def\Dsl{\,\raise.15ex\hbox{/}\mkern-13.5mu D} %this one can be subscripted
\def\dsl{\raise.15ex\hbox{/}\kern-.57em\partial}

 %pound sterling
\def\lspace{\ifx\answ\bigans{}\else\qquad\fi}
\def\lbspace{\ifx\answ\bigans{}\else\hskip-.2in\fi} % $$\lbspace...$$
\def\boxeqn#1{\vcenter{\vbox{\hrule\hbox{\vrule\kern3pt\vbox{\kern3pt
	\hbox{${\displaystyle #1}$}\kern3pt}\kern3pt\vrule}\hrule}}}
\def\mbox#1#2{\vcenter{\hrule \hbox{\vrule height#2in
		\kern#1in \vrule} \hrule}}  %e.g. \mbox{.1}{.1}
%	matters of taste
%\def\tilde{\widetilde} \def\bar{\overline} \def\hat{\widehat}
%
% some sample definitions
  %     curly letters

\def\darr#1{\raise1.5ex\hbox{$\leftrightarrow$}\mkern-16.5mu #1}
 %pound sterling

 %puts a small half in a displayed eqn
\def\roughly#1{\raise.3ex\hbox{$#1$\kern-.75em\lower1ex\hbox{$\sim$}}}

%%%%%%%%%%%%%%%%%%%%%%%%%%%%%%%%%%%%%%%%%%%%%%%%%%%%%%%%%%%%%%%%%%%%%%%%%%%%%%
\newcount\yearltd\yearltd=\year
%\writedefs
%\input \jobname.defs
%%%%%%%%%%%%%%%%%%%%%%%%%%%%%%%%%%%%%%%%%%%%%%%%%%%%%%%%%%%%%%%%%%%%%%%%%%%%%%
\overfullrule=0pt
\parindent 25pt
\tolerance=10000
%\sequentialequations
%\draftmode
%%%%%%%%%%%%%%%%%%%%%%%%%%%%%%%%%%%%%%%%%%%%%%%%%%%%%%%%%%%%%%%%%%%%%%%%%%%%%%
\font\ensX=msbm10
\font\ensVII=msbm7
\font\ensV=msbm5
\newfam\math
\textfont\math=\ensX \scriptfont\math=\ensVII \scriptscriptfont\math=\ensV
\def\ensemble{\fam\math\ensX}
%%%%%%%%%%%%%%%%%%%%%%%%%%%%%%%%%%%%%%%%%%%%%%%%%%%%%%%%%%%%%%%%%%%%%%%%%%%%%
\def\ZZ{{\ensemble Z}}

\def\F(#1,#2){\langle\psi(z_{#1})\psi(z_{#2})\rangle_s}
\def\FB(#1,#2){\langle\psi(z_{#1})\tilde\psi(\bar z_{#2})\rangle_s}
\def\FBB(#1,#2){\langle\tilde\psi(\bar z_{#1})\tilde\psi(\bar z_{#2})\rangle_s}
\def\b(#1,#2){\langle\partial x(z_{#1}) x(z_{#2})\rangle}
\def\bb(#1,#2){\langle\bar\partial x(z_{#1}) x(z_{#2})\rangle}
\def\B(#1,#2){\langle\partial x(z_{#1})\partial x(z_{#2})\rangle}
\def\BB(#1,#2){\langle\partial x(z_{#1})\bar\partial x(\bar z_{#2})\rangle}
\def\BBB(#1,#2){\langle\bar\partial x(\bar z_{#1})\bar\partial x(\bar z_{#2})\rangle}
%%%%%%%%%%%%%%%%%%%%%%%%%%%%%%%%%%%%%%%%%%%%%%%%%%%%%%%%%%%%%%%%%%%%%%%%%%%%
\def\simlt{\mathrel{\lower2.5pt\vbox{\lineskip=0pt\baselineskip=0pt
           \hbox{$<$}\hbox{$\sim$}}}}
\def\simgt{\mathrel{\lower2.5pt\vbox{\lineskip=0pt\baselineskip=0pt
           \hbox{$>$}\hbox{$\sim$}}}}
%%%%%%%%%%%%%%%%%%%%references start here%%%%%%%%%%%%%%%%%%%%%%%%%%%%%%%%%%%%

\lref\CorleyHG{S.~Corley, D.~A.~Lowe and S.~Ramgoolam, {\sl Einstein-Hilbert
action on the brane for the bulk graviton}, JHEP {\bf 0107}, 030 (2001), [arXiv:hep-th/0106067].
%%CITATION = HEP-TH 0106067;%%
}

\lref\KiritsisTA{E.~Kiritsis and C.~Kounnas, {\sl Infrared regularization of superstring theory and the one loop calculation of coupling constants},
Nucl.\ Phys.\ B {\bf 442}, 472 (1995), [arXiv:hep-th/9501020].
%%CITATION = HEP-TH 9501020;%%
}

\lref\BainFB{P.~Bain and M.~Berg, {\sl Effective action of matter fields in
four-dimensional string  orientifolds}, JHEP {\bf 0004}, 013 (2000), [arXiv:hep-th/0003185].
%%CITATION = HEP-TH 0003185;%%
}

\lref\VerlindeFY{H.~Verlinde, {\sl Holography and compactification},
Nucl.\ Phys.\ B {\bf 580}, 264 (2000), [arXiv:hep-th/9906182];\hfill\break
%%CITATION = HEP-TH 9906182;%%
S.~B.~Giddings, S.~Kachru and J.~Polchinski, {\sl Hierarchies
from fluxes in string compactifications}, [arXiv:hep-th/0105097].
%%CITATION = HEP-TH 0105097;%%
}

\lref\Green{O.~Bergman, M.R.~Gaberdiel and M.B.~Green, {\sl D-brane
interactions in type~IIB plane-wave background}, [arxiv:hep-th/0205183].
%%CITATION = HEP-TH 0205183;%%
}

\lref\kata{T.~Katayanagi, {\sl Modular Invariance of Strings on
PP-Waves with RR Flux}, [arxiv:hep-th/0206010].
%%CITATION = HEP-TH 0206010;%%
}

\lref\Itzykson{C.~Itzykson, H.~Saleur,
{\sl Two-Dimensional Field Theory Close to Criticality}, J. Stat. Phys. 
{\bf 48}, 449  (1986).
}

\lref\GrossAR{D.~J.~Gross and P.~F.~Mende, {\sl The High-Energy Behavior Of
String Scattering Amplitudes}, Phys.\ Lett.\ B {\bf 197}, 129 (1987)~;
%%CITATION = PHLTA,B197,129;%%
D.~J.~Gross and P.~F.~Mende, {\sl String Theory Beyond The
Planck Scale}, Nucl.\ Phys.\ B {\bf 303}, 407 (1988). 
%%CITATION = NUPHA,B303,407;%%
}

\lref\GrossIV{D.~J.~Gross and E.~Witten, {\sl Superstring Modifications Of
Einstein's Equations}, Nucl.\ Phys.\ B {\bf 277}, 1 (1986).
%%CITATION = NUPHA,B277,1;%%
}

\lref\KiritsisEM{E.~Kiritsis and B.~Pioline, {\sl On R**4 threshold corrections in type IIB string theory and (p,q) string  instantons},
Nucl.\ Phys.\ B {\bf 508}, 509 (1997)
[arXiv:hep-th/9707018].
%%CITATION = HEP-TH 9707018;%%
}

\lref\DGP{G.~R.~Dvali, G.~Gabadadze and M.~Porrati, {\sl 4D gravity on a brane in 5D Minkowski space},
Phys.\ Lett.\ B {\bf 485}, 208 (2000) [arXiv:hep-th/0005016].
%%CITATION = HEP-TH 0005016;%%
}

\lref\DG{G.~R.~Dvali and G.~Gabadadze, {\sl Gravity on a brane in
infinite-volume extra space}, Phys.\ Rev.\ D {\bf 63}, 065007 (2001) [arXiv:hep-th/0008054].
%%CITATION = HEP-TH 0008054;%%
}

\lref\TseytlinR{J.~G.~Russo and A.~A.~Tseytlin, {\sl One-loop
four-graviton amplitude in eleven-dimensional supergravity}, Nucl.\
Phys.\ B {\bf 508}, 245 (1997), [arXiv:hep-th/9707134].
%%CITATION = HEP-TH 9707134;%%
}

\lref\GreenGV{M.~B.~Green, M.~Gutperle and P.~Vanhove, {\sl One-loop
in Eleven Dimensions}, Phys.\ Lett.\ B {\bf 409}, 177 (1997),
[arXiv:hep-th/9706175].
%%CITATION = HEP-TH 9706145;%%
}

\lref\BGreenVanhove{M.~B.~Green and P.~Vanhove, 
{\sl D-instantons, strings and M-theory}, Phys.\ Lett.\ B {\bf 408}, 122
(1997), [arXiv:hep-th/9704145]. 
%%CITATION = HEP-TH 9704145;%% 
}
\lref\BGreenVanhoveB{M.~B.~Green, H.-K.~Kwon and P.~Vanhove,  {\sl Two
loops in eleven dimensions}, Phys.\ Rev.\ D {\bf 61}, 104010 (2000)
[arXiv:hep-th/9910055].
%%CITATION = HEP-TH 9910055;%%
}

\lref\GimonRQ{E.~G.~Gimon and J.~Polchinski, {\sl Consistency Conditions for
Orientifolds and D-Manifolds}, Phys.\ Rev.\ D {\bf 54}, 1667 (1996)
[arXiv:hep-th/9601038].
%%CITATION = HEP-TH 9601038;%%
}

\lref\ErlerAN{J.~Erler, D.~Jungnickel, J.~Lauer and J.~Mas, {\sl String emission from twisted sectors: cocycle operators and modular background symmetries}, Annals Phys.\  {\bf 217}, 318 (1992).
%%CITATION = APNYA,217,318;%%
}

\lref\GinspargUI{P.~Ginsparg, {\sl Applied Conformal Field Theory}, Les
Houches (1988).} 

\lref\GreenMN{ M.~B.~Green, J.~H.~Schwarz and E.~Witten, {\sl Superstring Theory}, Cambridge, (1987).}

\lref\CarloReview{C.~Angelantonj and A.~Sagnotti, {\sl Open Strings},
[arXiv:hep-th/0204189].
%%CITATION = HEP-TH 0204089;%%
}

\lref\BachasZT{C.~Bachas and C.~Fabre, {\sl Threshold Effects in
Open-String Theory}, Nucl.\ Phys.\ B {\bf 476}, 418 (1996),
[arXiv:hep-th/9605028]. 
%%CITATION = HEP-TH 9605028;%%
}

\lref\DixonQV{L.~J.~Dixon, D.~Friedan, E.~J.~Martinec and S.~H.~Shenker,
{\sl The Conformal Field Theory Of Orbifolds}, Nucl.\ Phys.\ B {\bf 282}, 13 (1987).
%%CITATION = NUPHA,B282,13;%%
}

\lref\HamidiVH{S.~Hamidi and C.~Vafa, {\sl Interactions On Orbifolds},
Nucl.\ Phys.\ B {\bf 279}, 465 (1987).
%%CITATION = NUPHA,B279,465;%%
}

\lref\AntoniadisB{I.~Antoniadis, C.~Bachas, C.~Fabre, H.~Partouche and
T.~R.~Taylor,  {\sl Aspects of type I - type II - heterotic triality
in four dimensions}, Nucl.\ Phys.\ B {\bf 489}, 160 (1997), [arXiv:hep-th/9608012]. 
%%CITATION = HEP-TH 9608012;%% 
}

\lref\AntoniadisPartoucheRTaylor{I.~Antoniadis, H.~Partouche and T.~R.~Taylor, 
{\sl Duality of N = 2 heterotic-type I compactifications in four  dimensions}, Nucl.\ Phys.\ B {\bf 499}, 29 (1997), [arXiv:hep-th/9703076]. 
%%CITATION = HEP-TH 9703076;%% 
}

\lref\AntoniadisFMN{I.~Antoniadis, S.~Ferrara, R.~Minasian
and K.~S.~Narain,  {\sl $R^4$ couplings in M- and type II theories on 
Calabi-Yau spaces}, Nucl.\ Phys.\ B {\bf 507}, 571 (1997), 
[arXiv:hep-th/9707013]. 
%%CITATION = HEP-TH 9707013;%% 
}

\lref\VafaWitten{C.~Vafa and E.~Witten, {\sl A One loop test of string
duality}, Nucl.\ Phys.\ B {\bf 447}, 261 (1995), [arXiv:hep-th/9505053].
%%CITATION = HEP-TH 9505053;%% 
} 

\lref\JDuffTLiuMinasian{M.~J.~Duff,
J.~T.~Liu and R.~Minasian, {\sl Eleven-dimensional origin of string /   
string duality: A one-loop test}, Nucl.\ Phys.\ B {\bf 452}, 261 (1995),
[arXiv:hep-th/9506126]. 
%%CITATION = HEP-TH 9506126;%% 
}

\lref\Ldim{I.~Antoniadis, {\sl A Possible New Dimension At A Few TeV},
Phys.\ Lett.\ B {\bf 246}, 377 (1990);\hfill\break
%%CITATION = PHLTA,B246,377;%%
I.~Antoniadis and K.~Benakli, {\sl Limits on extra dimensions in orbifold
compactifications of superstrings}, Phys.\ Lett.\ B {\bf 326}, 69 (1994),
[arXiv:hep-th/9310151];\hfill\break
%%CITATION = HEP-TH 9310151;%%
I.~Antoniadis, N.~Arkani-Hamed, S.~Dimopoulos and G.~R.~Dvali,
{\sl New dimensions at a millimeter to a Fermi and superstrings at a TeV},
Phys.\ Lett.\ B {\bf 436}, 257 (1998),
[arXiv:hep-ph/9804398].
%%CITATION = HEP-PH 9804398;%%
}

\lref\RS{L.~Randall and R.~Sundrum, {\sl An alternative to compactification},
Phys.\ Rev.\ Lett.\  {\bf 83}, 4690 (1999),
[arXiv:hep-th/9906064];\hfill\break
%%CITATION = HEP-TH 9906064;%%
L.~Randall and R.~Sundrum, {\sl A large mass hierarchy from a small extra
dimension}, Phys.\ Rev.\ Lett.\  {\bf 83}, 3370 (1999),
[arXiv:hep-ph/9905221].
%%CITATION = HEP-PH 9905221;%%
}

\lref\KTT{E.~Kiritsis, N.~Tetradis and T.~N.~Tomaras,
{\sl Induced brane gravity: realizations and limitations},
JHEP {\bf 0108}, 012 (2001), [arXiv:hep-th/0106050].
%%CITATION = HEP-TH 0106050;%%
}

\lref\Ka{Z.~Kakushadze, {\sl Orientiworld}, JHEP {\bf 0110}, 031 (2001),
[arXiv:hep-th/0109054].
%%CITATION = HEP-TH 0109054;%%
}

\lref\AGN{I.~Antoniadis, E.~Gava and K.~S.~Narain, 
{\sl Moduli corrections to gravitational couplings from string loops},
Phys.\ Lett.\ B {\bf 283}, 209 (1992), [arXiv:hep-th/9203071].
%%CITATION = HEP-TH 9203071;%%
}

\lref\ABD{I.~Antoniadis, C.~Bachas and E.~Dudas,
{\sl Gauge couplings in four-dimensional type I string orbifolds},
Nucl.\ Phys.\ B {\bf 560}, 93 (1999), [arXiv:hep-th/9906039].
%%CITATION = HEP-TH 9906039;%%
}

\lref\Grisaru{M.T.~Grisaru, A.E.M.~van de Ven and D.~Zanon,
Nucl.\ Phys.\ B {\bf 277}, 409 (1986).
%%CITATION = NUPHA,B277,388;%%
}

\lref\Kohlprath{ E.~Kohlprath, {\sl Renormalization of the Planck mass for
type II superstrings on symmetric orbifolds}, [arXiv:hep-th/0207023].
%%CITATION = HEP-TH 0207023;%%
}

\lref\AP{I.~Antoniadis and B.~Pioline, {\sl Low-scale closed strings and their
duals}, Nucl.\ Phys.\ B {\bf 550}, 41 (1999), [arXiv:hep-th/9902055];
%%CITATION = HEP-TH 9902055;%%
K.~Benakli and Y.~Oz,
{\sl Small instantons and weak scale string theory},
Phys.\ Lett.\ B {\bf 472}, 83 (2000), [arXiv:hep-th/9910090].
%%CITATION = HEP-TH 9910090;%%
}

\lref\ADG{I.~Antoniadis, S.~Dimopoulos and A.~Giveon,
{\sl Little string theory at a TeV}, JHEP {\bf 0105}, 055 (2001),
[arXiv:hep-th/0103033]. 
%%CITATION = HEP-TH 0103033;%%
}

\lref\CandelasNC{
P.~Candelas, D.~E.~Diaconescu, B.~Florea, D.~R.~Morrison and G.~Rajesh,
{\sl Codimension-three bundle singularities in F-theory},
JHEP {\bf 0206}, 014 (2002), [arXiv:hep-th/0009228].
%%CITATION = HEP-TH 0009228;%%
}

\lref\doph{E.~D'Hoker and D.~H.~Phong, {\sl The Geometry Of String
Perturbation Theory}, Rev.\ Mod.\ Phys.\  {\bf 60}, 917 (1988).
%%CITATION = RMPHA,60,917;%%
}

%%%%%%%%%%%%%%%%%%%%%%%%%%%%%%%%%%%%%%%%%%%%%%%%%%%%%%%%%%%%%%%%%%%%%%%%%%%%%%
\noblackbox
\baselineskip 14pt plus 2pt minus 2pt
\Title{\vbox{\baselineskip12pt
\hbox{hep-th/0209030}
\hbox{CPHT-RR 067.0902}
\hbox{CERN-TH/2002-220}
\hbox{SPHT-T02/035}
}}
{\vbox{
\centerline{ Non-compact Calabi--Yau Manifolds and Localized Gravity}
}}
\centerline{Ignatios Antoniadis$^{1,a,}$\footnote{${}^\dagger$}{\sevenrm On leave
of absence from CPHT {\'E}cole Polytechnique$^{2}$.}, Ruben Minasian$^{2,b}$ and 
Pierre Vanhove$^{1,3,c}$}
\medskip
\centerline{${}^1$ \sl CERN Theory Division CH-1211 Geneva 23, Switzerland}
\centerline{${}^2$ \sl CPHT {\'E}cole Polytechnique (UMR du CNRS 7644) 91128
Palaiseau Cedex,  France}
\centerline{${}^3$ \sl CEA/DSM/SPhT, Unit{\'e} de recherche associ{\'e}e au CNRS, CEA/Saclay, 91191 Gif-sur-Yvette, France}
\centerline{$^a${\tt ignatios.antoniadis@cern.ch}, 
$^b${\tt ruben@cpht.polytechnique.fr},
$^c${\tt vanhove@spht.saclay.cea.fr}}
\bigskip

%% abstract
\medskip
\centerline{{\bf Abstract}}

We study localization of gravity in flat space in superstring theory. We find
that an induced Einstein-Hilbert term can be generated only in four dimensions,
when the bulk is a non-compact Calabi--Yau threefold with non-vanishing Euler
number. The origin of this term is traced to $R^4$ couplings in ten
dimensions. Moreover, its size can be made much larger than the
ten-dimensional gravitational Planck scale by tuning the string coupling to be
very small or the Euler number to be very large. We also study the width of
the localization and discuss the problems for constructing realistic string
models with no compact extra dimensions.

%%%%%%%%%%%%%%%%%%%%%%%%%%%%%%%%%%%%%%%%%%%%%%%%%%%%%%%%%%%%%%%%%%%%%%%%%%%%%%
\noblackbox
\baselineskip 14pt plus 2pt minus 2pt
\Date{}
%%%%%%%%%%%%%%%%%%%%%%%%%%%%%%%%%%%%%%%%%%%%%%%%%%%%%%%%%%%%%%%%%%%%%%%%%%%%%
\newsec{Introduction}
 
String models with large extra dimensions have lead to many interesting
phenomenological developments in the recent years~\refs{\Ldim}.  In such 
models,  gauge degrees of freedom are usually localized on D-branes while 
gravity, described by closed strings, propagates in the bulk. Moreover, by 
tuning the size of the latter, one can achieve a large hierarchy between the 
string and four-dimensional (4d) Planck scales. In fact, the hierarchy problem 
is traded in this framework for the existence of large compact dimensions, 
with size much larger than the string length.

On the other hand, models with localized gravity have not yet a clear
realization in the context of string
theory~\refs{\RS,\DGP,\DG,\VerlindeFY,\KTT,\CorleyHG,\Ka}. Indeed, while largely inspired
by stringy developments and having used many string-theoretic techniques,
these models generally have not been seen as arising from string
theory. Moreover from stringy point of view they often appear to be rather ad
hoc.  Thus, string derivation of models with localized gravity not only
remains as one of the important problems, but may be helpful in establishing
the scales and other data that are required otherwise as an input. A
particularly attractive possibility is when the bulk is non compact which
avoids the problem of fixing the moduli associated to the size of the compactification manifold.

Since curved space is always difficult to handle in string theory, here we
concentrate on flat space with gravity localized on a subspace of the bulk
(for reasons that will become clear later, we intentionally avoid using 
the term ``brane"), and we shall demonstrate how it can be realized in string 
theory. The model was introduced in \refs{\DGP, \DG} and is based on
considering simultaneously Einstein-Hilbert (EH) actions in $D=4+n$ 
and $d=4$ dimensions:
\eqn\twoHE{
M^{2+n} \int d^{4+n}x\sqrt{g}{\cal R}_{(4+n)} + M_P^2 \int d^4x\sqrt{g}{\cal
R}_{(4)} \quad ;\quad M_P^2=M^{2+n}r_c^n\, , 
}
with $M$ and $M_P$ the respective Planck scales. Depending on the (possibly
independent) values of the two gravitational scales, a crossover parameter
controls the regime when the effective gravity is the lower- or
higher-dimensional one.

In the case of co-dimension one bulk ($n=1$) and $\delta$-function
localization, it is easy to see that the crossover scale is $r_c$. Indeed, for
distances smaller than $r_c$, the graviton propagator on the ``brane" exhibits
four-dimensional behavior with Planck constant $M_P$, while at large distances
it acquires a five-dimensional fall-off with Planck constant $M$. 
On the other hand, in the presence of non-zero brane
thickness $w$, a new crossover length-scale, $R_c=(wr_c)^{1/2}$, 
seems to appear. Below this scale the graviton propagator again becomes 
five-dimensional with an effective larger Planck constant 
$M_*=M(r_c/w)^{1/6}$~\refs{\KTT}.

The situation changes drastically for more than one non-compact bulk
dimensions, $n>1$, due to the ultraviolet properties of the higher-dimensional
theories. Thus, in the limit of zero thickness, the Newton's law is always
four-dimensional on the ``brane", while in the presence of a non-zero $w$
there is only one crossover length-scale, $R_c$:
\eqn\Rc{
R_c=w\left({r_c\over w}\right)^{n\over2}\, ,
}
above which one obtains a higher dimensional behavior~\refs{\KTT}. 

In this work, we study the possible stringy origin of two, one higher and one
lower dimensional, EH terms of the type~\twoHE, realizing the idea of
localized gravity. Clearly, string theory has a ten-dimensional EH term, and
thus, the question is how the lower-dimensional part arises, in how many
dimensions, and what are the parameters involved. A particularly important
point, which is necessary for the phenomenological viability of this scenario,
is that the strength of the lower (four) dimensional term, identified with the
Planck mass, must be much stronger than the higher dimensional gravitation
scale $M$.

It is well-known that the multi-graviton scattering in string theory can
generate higher-derivative couplings in curvature~\refs{\GreenMN}. These
generally go under the name of $R^4$ couplings and are well-studied in
ten-dimensional flat space $M_{10}$~\refs{\BGreenVanhove, \GreenGV,
\KiritsisEM, \TseytlinR, \BGreenVanhoveB}. So one may wonder if similarly, in
certain backgrounds of string theory, localized EH terms can be also
generated. As we will show, this indeed happens in type II superstring on $M_4
\times  X_6,$ where $M_4$ is the four-dimensional Minkowski space and $X_6$ is a
non-compact Ricci flat six-dimensional manifold with non-vanishing Euler
number (the compact case has been discussed in \refs{\AntoniadisFMN}).  We
will first derive the main features using a simple reasoning, based on the
analysis of the structure of the $R^4$ couplings, and then we will confirm the
results by performing explicit string computations.  An interesting aspect of
our analysis is that localization of the EH term is possible only in four
dimensions,\foot{Actually, in the strong coupling limit,
localization first occurs in five dimensions, as we will argue below by
lifting the effective action to M theory.} while in the non-compact case we
argue that it comes entirely from the type II closed string sector.

Our paper is organized as follows.  In section 2, we discuss the $R^4$
couplings and show how the localization arises in the context of the effective
field theory. We will also present the outline of our results. Sections 3 and
4 (and coresponding appendices~A and~B) are devoted to string calculations. In
section 3, we compute the corrections to the Planck mass from open strings and
show that they vanish in the decompactification limit (at least for
supersymmetric vacua). In section 4, we compute the corrections from closed
strings and we find a universal contribution localized in four dimensions and
proportional to the Euler number of the internal space. In section 5, we
analyze the width of the localized terms. Finally, in section 6, we discuss
the problems and the conditions for constructing realistic string models with
no compact extra dimensions and localized gravity.

%%%%%%%%%%%%%%%%%%%%%%%%%%%%%%%%%%%%%%%%%%%%%%%%%%%%%%%%%%%%%%%%%%%%%%%%%%%

\newsec{Low-energy effective action and outline of the results}

In string theory, corrections to the two-derivative EH action are in general
very restrictive. For instance, in the heterotic string, they vanish to all
orders in perturbation theory~\refs{\AGN}. On the other hand, in type II
theories, they are constant (moduli independent) and receive contributions
only from tree and one loop level (at least for supersymmetric
backgrounds)~\refs{\AntoniadisFMN}. In this work, we will show that they
actually describe localized terms in four dimensions, which therefore survive
in the non-compact (decompactification) limit. Finally, in type I theory,
there are moduli dependent corrections generated by open
strings~\refs{\AntoniadisB, \AntoniadisPartoucheRTaylor}, but as we will show
in section 3, they vanish in the decompactification limit.

Below, we describe the corrections in type II theories from the effective field
theory point of view. In ten dimensions, the type II effective action including
the tree-level and one-loop terms is given
by~\refs{\BGreenVanhove,\KiritsisEM,\BGreenVanhoveB,\AntoniadisFMN}:
$$\eqalign{
{1\over (2\pi)^7 l_s^8} \int_{M_{10}} %d^{10}x \sqrt{-g}  
& \left(e^{-2\phi} {\cal R}_{(10)} + {2\zeta(3)\over 3\cdot   2^7}l_s^6 e^{-2\phi}
(t_8t_8-{1\over 4} \epsilon_{8}\epsilon_{8}) R^4 %\cr 
+  {2\pi^2l_s^6\over 3^2\cdot     2^7}  
(t_8t_8 \pm   {1\over 4} \epsilon_{8}\epsilon_{8}) R^4 +\cdots \right),
}
$$
where $l_s=M_s^{-1}$ is the string length scale, $\phi$ is the dilaton field
determining the string coupling $g_s=e^{\langle\phi\rangle}$, and ${\cal
R}_{(10)}$ is the ten-dimensional curvature scalar. The rank 8 tensor $t_8$ is
defined  as $t_8M^4\equiv -6({\rm tr}M^2)^2 +24{\rm tr}M^4$,
$\epsilon_8$ is the rank 8 totally antisymmetric tensor, and the $+/-$ sign in
the last term corresponds to the type IIA/B theory. Here, we have dropped CP-odd
terms proportional to $t_8\epsilon_8$. Using differential form notations, we
can rewrite the above action as:
\eqn\etypeIIdf{\eqalign{
{1\over (2\pi)^7 l_s^8}& \int_{M_{10}} e^{-2\phi} {\cal R}_{(10)} +
{1\over 3(4\pi)^7 l_s^2}\int_{M_{10}}\,  \left(
2\zeta(3)e^{-2\phi} +  {2\pi^2\over 3}\right)t_8t_8 R^4 \cr
&-{1\over 3(4\pi)^7 l_s^2} \int_{M_{10}} \left( 2\zeta(3)e^{-2\phi}
\mp {2\pi^2\over 3}\right) R\wedge R \wedge R\wedge R \wedge e \wedge e
+\cdots
}}

As already mentioned, we are interested in considering these couplings on 
a six-dimensional non-compact Calabi--Yau (CY) manifold $X_6$, and thus take a 
background $M_4 \times X_6$. The discussion in this section is 
heuristic and is concerned mostly with analyzing the structure of the various
terms in \etypeIIdf. In particular, we ignore here issues connected with 
$X_6$ being non-compact and more importantly having boundaries. We will 
see in section 4 how these points get settled by string calculations.

Due to the fact that spacetime is a product of two manifolds, most of the terms 
in the $t_8t_8 R^4$ part drop out and at any rate  do not contribute to 
the Einstein-Hilbert action in four dimensions but to $R^2$
terms~\refs{\AntoniadisFMN}.  In principle, on a six-manifold a cosmological
constant $\int_{X_6} t_8t_8R^4$ could be induced, but for a CY threefold this 
contribution vanishes. Indeed, we recall that $\int_{X_6} (t_8t_8 -
{1\over8}\epsilon_8\epsilon_8) R^4=0$ due to the existence of a covariantly
constant spinor on Ricci-flat  K{\"a}hler background with SU(3)
holonomy~\refs{\GrossIV}. Since on six dimensional manifolds
$\int_{X_6}\epsilon_8\epsilon_8 R^4=0$, there is no contribution to the four dimensional cosmological constant from $t_8t_8R^4$.

Essentially we have to worry only about contributions from the ``geometric" 
part, $\int R^4\wedge e^2$.  Here the  discussion  closely
follows~\refs{\AntoniadisFMN}, and due to the ``internal" part being a  total
derivative the result is practically identical to the case where $X_6$ is compact: 
\eqn\eLocal{
{1\over (2\pi)^7 l_s^8}\int_{M_4 \times   X_6}\, e^{-2\phi}{\cal R}_{(10)} + 
{\chi\over 3(4\pi)^7 l_s^2}
 \int_{M_4}\, \left(-2\zeta(3)e^{-2\phi} 
\pm   4\zeta(2)\right) {\cal R}_{(4)} %\wedge e\wedge e 
\, .
}
A number of conclusions (confirmed by string calculations) can be reached 
by looking closely at \eLocal:

\item{$\triangleright$} Localization requires $X_6$ to have a non-zero Euler
characteristic $\chi\neq 0$. Actually, $\chi$ counts the difference between
the numbers of ${\cal N}=2$ vector multiplets and hypermultiplets:
$\chi=\pm 4(n_V-n_H)$ in type IIA/B (where the graviton multiplet counts as one
vector).\foot{Note that in the non-compact limit, the Euler number can in
general split in different singular points of the internal space, giving rise
to different localized terms. This is clear from the orbifold examples that we
discuss in more detail in section 4. To simplify our discussion, in the
following we consider the simplest case where $\chi$ is concentrated on one
singular point.}

\item{$\triangleright$} The structure of the localized ${\cal R}_{(4)}$ term
coming from the closed string sector is universal, independent of the
background geometry and dependent only on the internal topology.

\item{$\triangleright$} It is a matter of simple inspection to see that if one
wants to have a localized EH term in less than ten dimensions, namely
something linear in curvature, with non-compact internal space in all
directions, the only dimension where this is possible is four.

\noindent A brief comment on the last item. We are not attempting to
discuss here from first principles why a localized EH action should
be preferable to say $R^2$ gravity. Phenomenologically speaking, this
is clear enough. Here we simply pursue the goal of obtaining
localized gravity of EH-type, without a priori fixing the
number of dimensions, and we see that it is possible only in four.

The next question is to study the conditions for which the localized term
becomes much more important than the bulk (ten-dimensional) EH action.  It is
interesting that this is indeed the case in the weak coupling limit, where the
one-loop contribution in \eLocal~can be ignored and the relevant scale of the
localized four-dimensional (4d) term $M_s/g_s$ is much larger than the
corresponding 10d scale $M_s/g_s^{1/4}$. Moreover, the 4d contribution can be
further enhanced by a large Euler number $\chi$. Note that $\chi$ should be
negative in order to obtain the correct sign for the gravity kinetic
terms.\foot{The Planck mass receives opposite contributions from vector
  multiplets and hypermultiplets.}

What about the strong coupling limit? Since type IIB theory  remains invariant 
under S-duality, we do not expect to find anything new in this limit. In type IIA on 
the other hand, the strong coupling limit is taking us to M theory. By lifting the
action \etypeIIdf\ in eleven dimensions and converting everything to the M theory
frame, we get
\eqn\eMtheory{\eqalign{
{1\over 2(2\pi l_M)^9}& \int_{M_{11}} {\cal R}_{(11)} +{1\over
(4\pi)^8\cdot    3}\int_{M_{10}\times   S^1}   
\left( {2\zeta(3)\over R_{11}^3} + {4\zeta(2)\over l_M^3}\right)t_8t_8 R^4 \cr
&-{1\over (4\pi)^8\cdot    3} \int_{M_{10}\times   S^1} 
\left( {2\zeta(3)\over R_{11}^3}- {4\zeta(2)\over l_M^3}\right) 
R\wedge R \wedge R\wedge R\wedge e \wedge e \wedge e +\cdots
}}
where $l_M$ is the 11d Planck scale and $R_{11}$ is the radius of the eleventh
dimension with $M_{11}=M_{10}\times   S^1$. Considering now a background
$M_5\times   X_6,$ with $M_5 = M_4\times   S^1$ and $X_6$ a non-compact Calabi--Yau, and
taking the large radius limit $R_{11}\to\infty$ (string strong coupling), we
find the action:
\eqn\eLocalM{
{1\over 2(2\pi l_M)^9} \int_{M_5\times   X_6} {\cal R}_{(11)} + 
{\chi\over 36(4\pi)^5 l_M^3} \int_{M_4\times   S^1} {\cal R}_{(5)}\, .
}
As before, essentially we had to worry only for contributions of the
``geometric"  part, $\int R^4\wedge e^3$ (the other geometric piece from
eleven dimensions  $\int C_3\wedge X_8(R)$
\refs{\VafaWitten,\JDuffTLiuMinasian} is not important  for our purposes and
has been ignored). Thus, as a consequence of the power-dependence on $R_{11}$
in~\eMtheory, a simultaneous localization of gravity in four and
five dimensions is prevented. Weak coupling localizes gravity in four
dimensions, while we find a five-dimensional localization in type IIA strong
coupling limit. However, in this case, the strength of the 5d localized term
is given by the same scale as the eleven-dimensional one $l_M^{-1}$ and can be
enhanced only by considering large $\chi$. Therefore, in the following we will
concentrate on the ten-dimensional string theory case \eLocal, in the weak
coupling limit, and we come back in M theory only in the last section. 

In open superstring models one can think of diagrams that may in principle
induce an Einstein-Hilbert term in six dimensions with ${\cal N}=1$
supersymmetry. But in the next section we will show that the coefficient of
such a term vanishes by the tadpole cancellation condition, and thus no such
term is generated. This is a bit surprising since only massless
six-dimensional states contribute to the amplitudes~\refs{\AntoniadisB}.

%%%%%%%%%%%%%%%%%%%%%%%%%%%%%%%%%%%%%%%%%%%%%%%%%%%%%%%%%%%%%%%%%%%%%%%%%%%%%%
\newsec{Planck mass corrections in open string models}

As mentioned above, a simple inspection of worldsheets with boundaries
suggests that in type I theory one-loop corrections to the Planck mass could
be generated already in six dimensions. In fact, the moduli dependence of such
corrections in four dimensions were computed in \refs{\AntoniadisB,
\AntoniadisPartoucheRTaylor} for ${\cal N}=2$ supersymmetric compactifications
on $T^2\times  K3$ and were found to depend only on the $T^2$ moduli and to be
proportional to an index given by the difference of ${\cal N}=2$ vector
multiplets and hypermultiplets. In this section, we compute the three-graviton
amplitude in order to determine the complete form of the corrections and find
whether the result decompactifies to a finite non zero contribution in six
dimensions. The rules for the normalizations and contractions are the same as
in \refs{\AntoniadisB}.

Only the even-spin structure can contribute to the Planck mass corrections. At
the two derivative level, we have to consider the contractions involving four
fermions from the following insertions of three graviton vertex operators with
corresponding spacetime momenta $k_i$:
$$
\eqalign{
\hbox{\rm at position}\ z_1 & : (\partial_z x^\mu    + {i\over 2}k_1\cdot \psi
\psi^\mu   ) \,(\bar\partial_{\bar z} x^\nu   - {i\over 2}k_1\cdot \bar\psi
\bar\psi^\nu ) \, e^{ik_1\cdot x}:(z_1,\bar z_1)\cr
\hbox{\rm at position}\ z_2 &: ( \partial_z x^\mu    + {i\over 2}k_2\cdot \psi
\psi^\mu   ) \,(\bar\partial_{\bar z} x^\nu - {i\over 2}k_2\cdot \bar\psi
\bar\psi^\nu ) \, e^{ik_2\cdot x}:(z_2,\bar z_2)\cr
\hbox{\rm at position}\ z_3 &: ( \partial_z x^\mu    + {i\over 2}k_3\cdot \psi
\psi^\mu   ) \,(\bar\partial_{\bar z} x^\nu - {i\over 2}k_3\cdot \bar\psi
\bar\psi^\nu ) \, e^{ik_3\cdot x}:(z_3,\bar z_3)
}$$
where $x$ are the spacetime coordinates while $\psi$ and $\bar\psi$ their left
and right 2d fermionic superpartners, depending on the worldsheet positions
$z_i$ and $\bar z_i$. The possible contractions are:
\eqn\eContractionseven{
\eqalign{
A &=\F(1,2)^2 \BB(3,2) \BBB(1,3)\cr
& - \FB(1,2)^2 \B(3,2) \BBB(3,1) + c.c. \cr
B&=\F(1,2)^2 \BB(3,1) \BBB(2,3)\cr
& - \FB(1,2)^2 \BB(3,1) \BB(2,3) + c.c.\cr
}}
with a corresponding tensorial structure $T_{A/B}$ in matrix notation. After
summation over all permutations of the external states, one finds:
\eqn\eTensor{
T_A =T_B =k_1\cdot    k_2\, tr(\zeta_1\zeta_3\zeta_2) - 
k_1\cdot    \zeta_2\cdot    \zeta_3\cdot   \zeta_1\cdot   k_2 + perm\, ,
}
with $\zeta_i^{\mu   \nu}$ being the graviton polarization tensors. This is the
linearization of the kinetic term of the graviton ${\cal R}_{(4)}$.

 We remind the reader that in order to compute the amplitude correctly, the 
zero mode (lattice) parts need to be expressed in terms of the open string 
proper-time  $t$, while the quantum fluctuations are functions of the closed 
string  proper-time $\tau$. This is a consequence of computing the 
correlators on the double covering of the surface. 
Skipping the tensorial structure (see appendix~A for notations and 
conventions), we can present  any of the three amplitudes of interest  
($\sigma={\cal A}, {\cal M}, {\cal K}$ for the annulus, M{\"o}bius strip and Klein
bottle, respectively) in the form 
$$\eqalign{
&A_\sigma=-{2^9\over(4\pi)^2}\,   \int d^2z_1 d^2z_2
d^2z_3\int_0^\infty {dt\over t^3} {P^{(2)}(t)\over \eta^2(t)} \times \sum_{s=2,3,4}
(-1)^{s-1} {\theta_s^2(\tau)\over \eta^4(\tau)} \, {\rm Z}^{int}_{s,\sigma}(\tau)\cr
\times   &( \F(1,2)^2 \BB(3,2) \BBB(1,3) \cr
&- \FB(1,2)^2 \B(3,2) \BBB(3,1) + c.c.)\, ,
}$$
where $P^{(2)}(t)/\eta^2(t)$ is the $T^2$ momentum partition function,
${\rm Z}^{int}_{s,\sigma}$ is the contribution of the internal ${\cal N}=4$
superconformal theory describing $K3$, and $\theta_s^2/\eta^4$ is the
oscillator contribution of the bosonic and fermionic coordinates of ${\cal
M}^4\times   T^2$ dependent on the even spin structures $s$. 

Using the following identity valid for $K3$ models~\refs{\AntoniadisB}
\eqn\eSumSpin{
{1\over2}\sum_{s=2,3,4} (-1)^{s-1}\theta_s^2(\tau|u) {\rm Z}^{int}_{s,\sigma}
=  \theta_1^2(\tau|u)\times {\rm Z}^{int}_{s=1,\sigma}\ ,
}
the even spin structure sum and the fermionic contractions yield:
\eqn\eAs{\eqalign{
A_\sigma&=-{2^6\over\pi^2} \, {\rm Z}^{int}_{s=1,\sigma}
\int_0^\infty {dt\over t^3} P^{(2)}(t)\times   \int d^2z_1 d^2z_2 d^2z_3 \cr
&\times   (\BB(3,2) \BBB(1,3) -  \B(3,2) \BBB(3,1) + c.c.) \, .
}}
Note that ${\rm Z}^{int}_{s=1,\sigma}$ is an index in the odd spin structure
$s=1$, $t$-independent, determined from the massless content of the theory in
six dimensions. Using now the identity (5.7) of~\refs{\AntoniadisB}, that for
a periodic function on the covering torus ${\cal T}$ of the open surface 
$$
\int_\sigma \partial_w f(w) - \partial_w f(I_\sigma(w))= \int_{\cal T}
\partial_w f(w) =0\, ,
$$
with $I_\sigma$ the corresponding $\ZZ_2$ involution, the amplitude reduces to
\eqn\eFinalA{\eqalign{
A_\sigma&=-{2^7\over\pi^2} \times   {\rm Z}^{int}_{s=1,\sigma}
\int_0^\infty {dt\over t^3}
P^{(2)}(t) \int d^2z_1 d^2z_2 d^2z_3 \times   \left(\pi\over 4\tau_2\right)^2\cr 
&=- {\rm Z}^{int}_{\sigma}
\int_0^\infty {dt\over t^3}\times   \tau_2 \eta^4(\tau)
\times   {P^{(2)}(t)\over \eta^4(\tau)}\, .\cr
}}
Here, we used that in terms of the closed string proper-time
$\tau=\tau_1+i\tau_2$, all diagrams $\sigma$ have the same volume $\tau_2/2$.

Next, we convert the amplitude to the transverse closed string channel with the
following sequence of changes of 
variables for each diagram (see \refs{\CarloReview,\GimonRQ,\BachasZT} and Appendix~A):
\eqn\eOtoC{\eqalign{
\sigma={\cal K}:& \quad \tau=2 it  \to -{1\over\tau} = i\ell \cr
\sigma={\cal A}:& \quad \tau={it\over2}  \to -{1\over\tau} = i\ell\cr
\sigma={\cal M}:& \quad \tau={1+it\over2} \to -{1\over\tau} \to -{1\over\tau}+2 \to
\left({1\over\tau}-2\right)^{-1}= -{1\over2}+{i\over2t} = i\ell -{1\over2}
}}
The one-loop correction to the Planck mass takes the form
\eqn\edeltaI{
\delta= -T_2\, \int_0^\infty d\ell \ell
\left[{1\over 8} W^{(2)}\left(\ell\over2\right) {\rm Z}^{int}_{\cal A}+ 2  
W^{(2)}\left(\ell\over2\right) {\rm Z}^{int}_{\cal M}+8 W^{(2)}(2\ell) Z^{int}_{\cal
K}\right] }
where $T_2$ is the volume of the two-torus, $W^{(2)}$ is the winding sum of
$T^2$ (obtained by Poisson resuming $P^{(2)}(t)$), and it is understood that
the internal partition function is restricted to the odd spin structure $s=1$
(which we dropped for simplicity) and thus $Z^{int}_{\cal A,M,K}$ are numbers.
The divergence for $\ell\to\infty$ is given by
\eqn\edeltaTwo{
\delta_{divergent}=-T_2 \,
\left[{1\over 8} {\rm Z}^{int}_{\cal A} + 2  {\rm Z}^{int}_{\cal M}+8
{\rm Z}^{int}_{\cal K} \right]\ \int^\infty d\ell\, \ell
}
where we used that in the $\ell\to\infty$ limit $W^{(2)}\to 1$. In appendix~A
we explain how the tadpole cancellation
conditions~\refs{\CarloReview, \GimonRQ,\BachasZT} imply that the prefactor
of the divergence vanishes   and the amplitude is finite.

On the other hand, the decompactification limit of \eFinalA\ to six dimensions
is  given by precisely the same expression~\edeltaTwo.
Therefore, there are no corrections to the
six-dimensional Planck mass consistently with the interpretation that all
contributions come from the localized fixed points.\foot{Strictly
speaking, in the non-compact case, tadpole cancellation is not required.
However, here we define non-compact spaces as decompactification limits
of compact ones. In the orbifold limits of $K3$ only twisted states
contribute to the correction to the Planck mass, for which the tadpole
condition has to be imposed even in the non-compact cases.} It
follows that the total correction to the EH action is given only by the four
dimensional contribution, whose $U$-dependence was analyzed in~\refs{\AntoniadisB}: 
\eqn\edeltaI{
\delta=-  {Z_2(U,\bar U)\over T_2}\,  \left[ {1\over 2}
{\rm Z}^{int}_{\cal A} +{1\over 2} {\rm Z}^{int}_{\cal M} 
+ 2{\rm Z}^{int}_{\cal K}\right]\ ,
}
where $U=U_1+iU_2$ is the complex structure (shape) modulus of $T^2$, and the
function $Z_2$ is given by:
$$
Z_2(U,\bar U) = \sum_{(m,n)\neq (0,0)} \, {U_2^2\over |m+nU|^4}=2\zeta(4)
U_2^2  + \pi\zeta(3)  U_2^{-1} + {\cal O}(\exp(-U_2))\ .
$$

Note that in ref.~\refs{\AntoniadisB} only the derivative with respect to the
$U$-modulus of the correction to the Planck mass was obtained, namely:
\eqn\edeltader{
4 U_2^2 \partial_U \partial_{\bar U} \delta = - 2 \,
{Z_2(U,\bar U)\over T_2}\,  \left[ {1\over 2} {\rm Z}^{int}_{\cal A} 
+{1\over 2} {\rm Z}^{int}_{\cal M} + 2{\rm Z}^{int}_{\cal K}\right]\, .
}
This relation follows trivially from~\edeltaI\ using the property that
the non-holomorphic Eiseinstein series $Z_s(U,\bar U)$ is an eigenfunction
of the $Sl(2,\ZZ)$-Laplacian with eigenvalue $s(s-1)$. Thus, our analysis
completely fixed the arbitrariness of possible zero modes of the Laplacian,
since we showed that the integration constant ($U$-independent piece) of
obtaining~\edeltaI\ from~\edeltader\ is vanishing after imposing the global
tadpole cancellation condition.

The square bracket of~\edeltaI\ contains the information about the
matter content of the $K3$-model, that we will denote $\chi_I$ since
this quantity counts the number of ${\cal N}=2$ vector multiplets minus the
number of hypermultiplets, in analogy with the Euler number $\chi$ of the
Calabi--Yau manifold in type IIA compactifications.

One may also study the five dimensional limit by considering a square $T^2$
torus with radii $R_{1,2}$ and taking the limit $R_1\to\infty$. In this case
$T_2=R_1R_2$ and $U=iR_1/R_2$. Expanding the modular form $Z_2(U,\bar U)$, the
Planck mass correction $\delta$ becomes:
\eqn\eExp{\eqalign{
\delta 
&=-{\pi\over 2}\, \chi_I \, R_1\times   \left(2\zeta(3) {1\over R_1^3}+
4\zeta(2){ \pi\over15}  {1\over R_2^3}   +  {\cal O}(\exp(-U_2))\right)
}}
and leads to a localized term in five dimensions:
$$
\sim -\chi_I\times {1\over R_2^3}\,\int d^5x\, \sqrt{g^{(5)}}\, {\cal R}_{(5)}\, .
$$
It is interesting to observe that this result reproduces the one found in the
M theory context~\eLocalM, upon the identification $R_{11}=R_1$ and
$l_M=R_2$. At the same time, the first term proportional to $\zeta(3)$
in~\eExp\ reproduces the subleading contribution proportional to $1/R_{11}^3$
in the second line of eq.~\eMtheory. Notice also the relative positive sign
between the two contributions of~\eExp, since we are in type~I
compactifications, corresponding to the type IIB choice of~\eLocal.\foot{A
more detailed analysis is however needed to understand the precise numerical
factors entering in this identification.}

One can also ask the question whether a localized EH term can be generated
already at the disk level, in analogy to the gauge kinetic terms. In fact, it
is known that gauge couplings in orientifold models are in general given (to
lowest order) as linear combinations of the dilaton $e^{-\phi}$ and the
various twisted moduli $m$ (blowing up modes), which are closed string
excitations localized at the orbifold fixed points: $1/g_a^2=e^{-\phi}+s_a m$,
with $s_a$ calculable constants~\refs{\ABD,\BainFB}. In the decompactification
limit, $e^{-\phi}$ leads to the usual 10d kinetic terms (actually
$(p+1)$-dimensional for a $p$-brane), while the terms proportional to $m$
yield additional contributions localized, say, at the origin of the internal
space. These terms can be computed by studying the one-loop infrared
divergence of the annulus amplitude in the closed string channel, generated by
the propagation of the massless twisted states. Following this method for
gravity, one should look for one-loop infrared divergences in the closed
string channel of an amplitude involving at least two gravitons with two
spacetime momenta each. A simple inspection of such amplitudes, analyzed for
instance in~\refs{\AntoniadisPartoucheRTaylor}, shows however the absence of
power divergences, at least for ${\cal N}=2$ supersymmetric compactifications.
It follows that there are no perturbative open string contributions that give
rise to localized graviton kinetic terms in six dimensions.

%%%%%%%%%%%%%%%%%%%%%%%%%%%%%%%%%%%%%%%%%%%%%%%%%%%%%%%%%%%%%%%%%%%%%%%%%%%%%%
\newsec{Planck mass corrections in type~II orbifolds models}

We return now to the contributions of closed type II strings propagating in
non-compact Calabi--Yau (CY) threefolds, that were outlined in section 2, in
the context of the effective field theory. As we pointed out, the localized
term becomes dominant at the weak coupling limit; it is determined at the
string tree-level from the four-loop beta-function of the two-dimensional
sigma model~\refs{\Grisaru}, and is proportional to the Euler number of the CY
manifold.

The aim of this and the following section is to study in detail the
localization properties of the EH term in the context of string theory, and
derive in particular the relevant width which determines the crossover
scale~\Rc. Since it is difficult to work in a generic Calabi--Yau space, even
at the string tree-level and in the non-compact limit, we would like to treat
a simple example, such as the orbifold case. However, in the orbifold limit,
the tree-level contribution to the localized EH action vanishes. This is easy
to see by inspection, for instance, of the K{\"a}hler metric of the untwisted
${\cal N}=2$ vector multiplets that should receive a perturbative 2d
$\sigma$-model correction proportional to $\zeta(3)\chi$. Such a correction is
though absent for orbifolds, since the tree-level metric of untwisted fields
can be exactly determined by truncation of the dimensionally reduced action
from ten dimensions, and can also be verified by a direct string computation
of the 4-point amplitude on the sphere.

For the above reason, in the following, we will restrict our analysis to the
one-loop correction and we will show that in the limit where the localized
gravity kinetic terms become dominant, their width is fixed by their strength,
which is in fact the only natural scale in this limit, as it has been also
argued in the literature before~\refs{\DG, \KTT}.  We will thus work in the
context of type~IIB theory compactified on the orbifold Calabi--Yau space
$CY^{(n_V,n_H)}=T^6/\ZZ_N$ with $n_V$ vector multiplets and $n_H$
hypermultiplets.  We will be particularly interested in the
decompactification limit when all the internal radii are sent to infinity.

The partition function of the model decomposes into a sum over three sectors
preserving different amount of supersymmetry: the ${\cal N}=(4,4)$ sector
where none of the coordinates is twisted, the ${\cal N}=(2,2)$ sector with two
untwisted internal coordinates, and the ${\cal N}=(1,1)$ sector where all the
Calabi--Yau coordinates are twisted:
$$
{\cal Z} = {\cal Z}^{(4,4)} + {\cal Z}^{(2,2)} + {\cal Z}^{(1,1)}\, .
$$
Following the computation of ref.~\refs{\AntoniadisFMN},\foot{One-loop Planck
mass corrections for type II $\ZZ_2$-orbifold was obtained
in~\refs{\KiritsisTA} by the background field method.}  only the odd-odd spin
structure contributes to the two derivative graviton kinetic terms. Moreover, 
only the ${\cal N}=(1,1)$ sector without internal fermionic zero modes gives
non-vanishing result. Since in the odd spin structure, the partition functions
of twisted fermions and bosons cancel among each other, we are left
with a constant proportional to the multiplicity of twisted states:
\eqn\eZodd{
{\cal Z}_{odd}^{int}={\cal Z}_{odd}^{(1,1)}=\sum_{f=0,\dots, n_f}\chi_f =\chi\, ,
}
where $f=0,\dots,n_f$ labels the fixed points and $\chi_f$ is the corresponding
contribution to the Euler number.
We consider now the one-loop amplitude involving two graviton zero modes with a
possible emission of a Kaluza-Klein (KK) excitation along the internal
directions. The vertex operator for the emission of an untwisted state,
with compact momentum $p$ and winding $\omega$, is
given in the $(0,0)$-ghost picture by the usual vertex operator
expressed in terms of the corresponding twisted coordinates $x_L$ and $x_R$
\refs{\HamidiVH}:
\eqn\eVTTU{\eqalign{
V_{(p,\omega)} &= :e^{i p_L\cdot x_L(z) + i p_R\cdot x_R(\bar z)}:\cr
V^{inv}_{(p,\omega)} &={1\over N}\, \sum_{k=0}^{N-1} V_{\gamma^k\cdot (p,\omega)}\,
, }}
where $p_L=(p+\omega)/\sqrt2$ and $p_R=(p-\omega)/\sqrt2$ are the internal left and
right momenta of the Narain lattice along the orbifold directions \refs{\HamidiVH}.
The invariant vertex operator is obtained after summation over its images
under the representation $\gamma$ of the action of the
orbifold group (normalized to the identity operator for
$V^{inv}_{(0,0)}$).\foot{See appendix~B for more details on the construction of
this vertex operator for $\ZZ_N$-orbifolds and computation of the twist
correlator.}

In the odd-odd spin structure, we will need to take one graviton vertex
in the $(-1,-1)$-ghost picture:
$$
V_{(-1,-1)}=\zeta_{MN}~:\psi^M
\bar\psi^N e^{-\varphi-\bar\varphi} \ V_{(p,\omega)}^{inv}\, e^{i k\cdot x}:\ ,
$$
where $\varphi$ and $\bar\varphi$ are the 2d superghosts.
The other vertices are in the $(0,0)$-ghost picture:
$$
V_{(0,0)}=\zeta_{MN}~:(\partial x^M+{i\over2}k\cdot \psi \psi^M)(\bar\partial
x^N-{i\over2}k\cdot \bar\psi \bar\psi^N) V_{(p,\omega)}^{inv}\, e^{i k\cdot x}:\ ,
$$ 
where the indices $M=(\mu ,I)$ can lie along the (non-compact) 4d directions $\mu $
and the (compact) 6d internal directions $I$. Besides the graviton vertices, one has
to consider insertions  of the holomorphic and anti-holomorphic world-sheet
supercurrents, $T_F= \partial x_\mu \psi^\mu + G_{IJ} \partial x^I \psi^J$ and
$\bar T_F= \bar\partial x_\mu \bar\psi^\mu + G_{IJ} \bar\partial x^I \bar\psi^J$,
respectively.

As mentioned above, the order ${\cal O}(k^2)$ gets contributions only
from the odd-odd spin-structure of the ${\cal N}=(1,1)$-twisted sector
of the three-graviton amplitude~\refs{\AntoniadisFMN}. The even-even
spin-structure and the other sectors with more  supersymmetry start
contributing from the order ${\cal O}(k^4)$.  At the order ${\cal O}(k^2)$,
after soaking the four spacetime fermionic zero-modes $\psi^{\mu =0,\dots, 3}$,
and taking into account that the contribution from the CY part
reduces to the twisted partition function ${\cal Z}^{(1,1)}_{odd}$ in
the odd-odd spin structure, which is the index~\eZodd, one finds:
\eqn\eThreeGrav{
\left\langle (V_{(0,0)})^3 \right\rangle ={\cal R}\,  \chi\, \int_{\cal F}
{d^2\tau\over\tau_2^2} + {\cal O}(k^4)\, ,
}
where the integration over $\tau$ is restricted in the fundamental domain
$\cal F$ for $Sl(2,\ZZ)$, and the linearized tensorial structure ${\cal R}$ is given by:
$$
{\cal R}= \varepsilon^{\mu _1\mu _2\mu _3\mu _4}
\varepsilon^{\nu_1\nu_2\nu_3\nu_4} k^1_{\mu _1} k^2_{\nu_1}
\zeta^1_{\mu _2\nu_2} \zeta^2_{\mu _3\nu_3}  \zeta^3_{\mu _4\nu_4} 
+{\rm permutations} \, .
$$
Note however that in the decompactification limit of the orbifold, the resulting
localized term at a given fixed point $f$ is obtained by replacing $\chi$
in~\eThreeGrav\ with the corresponding contribution $\chi_f$ defined in~\eZodd.

\bigskip
We turn now to the amplitude involving one Kaluza-Klein (KK) excitation of the
graviton with KK momentum $q$ and zero winding. Picking up the zero modes of the
fermions along the non-compact directions from $V_{(0,0)}(z_1)$ and
$V_{(-1,-1)}(z_2)$ and the zero mode parts from the contractions $\langle\partial
x^{\mu _2}(w)\bar\partial x^{\nu_3}(\bar z_3)\rangle$ and $\langle\bar\partial
x^{\nu_2}(\bar w)\partial x^{\mu _3}(z_3)\rangle$ between the supercurrents and
the vertex operator $V_{(0,0)}(z_3)$, we get using momentum conservation and the
mass-shell conditions $\sum_i k_i=0$, $(k_1)^2=(k_2)^2=0$ and $(k_3)^2 =-q^2$:
\eqn\eKKI{\eqalign{
 \left\langle (V_{(0,0)})^2 V_{(-1,-1)} \right\rangle^{(1)} &= {\cal R}
 \int_{\cal F} {d^2\tau\over
 \tau_2^2}\int \prod_{1\leq i\leq 3}{d^2z_i\over \tau_2}\, \left\langle\prod_{1\leq l\leq
3}:e^{ik_l\cdot  x}:\right\rangle\, {1\over N}{\sum_{(h,g)}}^\prime
 \left\langle V^{inv}_{(q,0)} \right\rangle^{(h,g)} \cr
&={\cal R}\,   \int_{\cal F} {d^2\tau\over \tau_2^2}\int \prod_{1\leq i\leq 3}{d^2z_i\over \tau_2} \ 
\prod_{1\leq i<j\leq 3}\chi_{ij}^{4\alpha^\prime k_i\cdot    k_j}\,  
{1\over N}{\sum_{(h,g)}}^\prime \langle
V^{inv}_{(q,0)}\rangle^{(h,g)} \cr 
&={\cal R}\, {1\over N^2}\sum_{f=0\cdots n_f\atop k=0\cdots N-1} e^{i\gamma^k
\, q\cdot x_f}  \int_{\cal F} {d^2\tau\over \tau_2^2}\int \prod_{1\leq i\leq
3}{d^2z_i\over \tau_2}\, {\sum_{(h,g)}}^\prime\, e^{\alpha^\prime  q^2
F_{(h,g)}(\tau,z_i)} \cr  }}
where $(h,g)=(l,m)\times v/N$ label the twisted
boundary conditions, with $l,m=0,\dots, N-1$;  $v/N$ is one of the three angles
of the orbifold action with $v$ being an integer between $1$ and $N-1$, usually
restricted by the crystallographic action and the supersymmetry restriction.
The prime in the sum excludes the untwisted sector $(h,g)=(0,0)$, which does
not contribute to the amplitude because of the zero modes.
In the second line of~\eKKI, we performed the bosonic contractions on the torus
defining as usual~\refs{\GreenMN}:
\eqn\eProp{
\chi_{ij} =2\pi \, e^{-\pi {(\Im{\rm m}(z_i-z_j))^2\over \tau_2}} \left|
\theta_1(z_i-z_j|\tau)\over \theta_1^\prime(0|\tau)\right|\, .
}
In the last line of~\eKKI, using theta-function identities, we
included the twisted sector contribution into the function
\eqn\eFs{
F_{(h,g)}(\tau,z_i)= -\ln(\chi_{12})^2 +\ln(\chi_{13})^2 +
\ln(\chi_{23})^2-\ln\Delta_h\ , }
where $\Delta_h$ is the coupling constant between twisted and untwisted states on the torus:
\eqn\eDeltaN{
\ln\Delta_h = 2\Psi(1)-\Psi(h)-\Psi(1-h)+ \sum_{n\in\ZZ\atop m>0}
{1\over|n+h|}\, e^{2i\pi m(g+(n+h)\tau_1) -2\pi m \tau_2 |n+h|}+c.c.
}
It depends on the modular parameter $\tau$ and is derived in
Appendix~B, as well as its modular properties, $\ln\Delta_h(-1/\tau)=
\ln\Delta_g(\tau)- \ln|\tau|^2$. In
the field theory limit, $\tau_2\to\infty$, the above expression reduces to the
standard tree-level coupling $\delta_h^{-\alpha^\prime q^2}$, with
$\ln\delta_h=2\Psi(1)-\Psi(h)-\Psi(1-h)$, between two twisted and one untwisted
state~\refs{\HamidiVH,\DixonQV}.

\bigskip
A second contribution to the amplitude comes by picking up the two fermionic
zero modes from $V_{(0,0)}(z_1)$, one from $V_{(0,0)}(z_3)$ and another from
$V_{(-1,-1)}(z_2)$. After contracting two internal fermions on each side (left and
right movers) and taking the bosonic zero modes from the $\langle\partial x
\bar\partial x\rangle$ contraction between the supercurrents, we find
$$
\eqalign{
 \left\langle \left(V_{(0,0)}\right)^2 V_{(-1,-1)} \right\rangle^{(2)}& = {\cal R}
 \int_{\cal F} {d^2\tau\over  \tau_2^2}\int \prod_{1\leq i\leq 3}{d^2z_i\over
\tau_2}\left\langle \prod_{1\leq l\leq 3}:e^{ik_l\cdot x}:\right\rangle\,
\cr
 \times {1\over N}\sum_{(h,g)}  \tau_2 & k^3_L k^1_{L'} G_{IJ} G_{I'J'}
\left\langle \partial x^I(w)
\bar\partial x^{I'}(\bar w)\right\rangle^{(h,g)} \left\langle\psi^J
\psi^L\right\rangle^{(h,g)}  \left\langle\bar\psi^{J^\prime}
\bar\psi^{L^\prime}\right\rangle^{(h,g)}\cr
&=0  \ .}
$$
The above expression vanishes because in the twisted sector bosonic
coordinates do not have zero modes: $\left\langle \partial x^I(w)
\bar\partial x^{I^\prime}(\bar w)\right\rangle^{(h,g)}=0$. Therefore, we are
left over only with the contribution~\eKKI.

%%%%%%%%%%%%%%%%%%%%%%%%%%%%%%%%%%%%%%%%%%%%%%%%%%%%%%%%%%%%%%%%%%%%%%%%%%%%%%
\newsec{Analysis of the width}

In order to extract the information on the effective width of the
localized term at the orbifold fixed point, we Fourier transform the
amplitude with respect to the KK momentum $q$ (in the Euclidean region $q^2<0$)
in all internal directions and take the decompactification limit of all radii
$R\to\infty$. Sitting at a fixed point of the orbifold, for instance the origin,
all other fixed points go to infinity and we are left
over with a Gaussian profile for the Planck mass correction $\delta$ in the 6d
internal position space $y$:
\eqn\eFZ{
\delta(y) ={1\over N}\, \int_{\cal F} {d^2\tau\over
 \tau_2^2}\int \prod_{1\leq i\leq 3}{d^2z_i\over \tau_2}\, \sum_{(h,g)}
{1\over (F_{(h,g)}(\tau,z_i))^3} \, e^{-{y^2\over \alpha^\prime F_{(h,g)}(\tau,z_i)}} \, .
}
Here, the sum over $k$ in~\eKKI\ cancels one factor of $N$. From this
expression we extract a form factor with a width $w$ associated with the fixed
point, and the corresponding localized induced EH effective action reads:
\eqn\echi{
{\chi_0\over l_s^2}\int d^4x d^6y\, \sqrt{g}\, {\delta}_w(y)  \, {\cal R}\, ,
}
where we have defined $\delta(y)\equiv\chi_0 M_s^2{\delta}_w(y)$.
The width $w$ of the form factor acts as a UV cutoff for the modes of the 4d
graviton propagating in the bulk~\refs{ \DG,\KTT}.

Indeed, the one-loop correction to the Planck mass $\delta(q)$ modifies the
Laplace equation for the Green's function as~\refs{\DGP, \DG}:
\eqn\eEqD{
M^{2+n} (k^2 + q^2)G(k,q) + M_P^2 k^2 \delta_w\star G(k,q) = 1\, ,
}
where the star stands for a convolution integral in a self-explanatory notation.
In the limit of vanishing width where $\delta_w(q)=1$, one can
partially Fourier transform from $q$ to $y$ and sit at the origin to find:
\eqn\eprop{
G(k,y=0)\sim{D(k,0)\over 1+r_c^6 k^2 D(k,0)}\, ,\qquad
D(k,0)=\int{d^6 q\over k^2+q^2}\, ,
}
where we utilized~\twoHE\ and for notational simplicity we use the same symbol for
a function and its Fourier transform. The above expression is of course formal,
since the bulk propagator $D$ has a short distance singularity at $y=0$, or at
$q\to\infty$. A finite width $w$ regulates the singularity and leads to the
crossover scale~\Rc\ for $n=6$, upon replacing $D(k,0)\sim w^{-4}$.

In order to determine in our case the effective width $w$ of the localized
gravitational kinetic term, we have to examine more closely the exponent of
the Gaussian profile~\eFZ. In our example, it is obviously fixed trivially by
the string length, which is the only available scale, times a numerical
constant. An additional parameter can however be introduced by varying the
Euler number $\chi$.  Thus, the one-loop induced four-dimensional Planck mass
in eq.~\twoHE\ becomes $M_P^2\sim\chi M_s^2$, while the ten-dimensional bulk
gravitational scale is $M^8\sim M_s^8/g_s^2$.

Using dimensional analysis in the limit $M_P\to\infty$, we expect the effective
width to vanish as a power of $l_P\equiv M_P^{-1}$: $w\sim
l_P^\nu/l_s^{\nu-1}$ with $\nu>0$. In refs.~\refs{\DG, \KTT}, it was argued
that $\nu=1$ and thus $w\sim\l_P$, which is the only left-over scale in the
decoupling limit of the effective field theory on the ``brane". However, we do
not see any a-priori reason for this argument to be valid in the context of
string theory which contains a finite fundamental length $l_s$. We thus allow
for a general positive exponent $\nu>0$, that we are going to determine from
the dependence of $w$ on $\chi$ in the limit $\chi\to\infty$.

To analyze the dependence of the exponent in~\eFZ\ on the Euler number, we should
study generic orbifolds. To simplify the discussion, we will
consider $\ZZ_N$ orbifolds with $N$ prime and take the limit of $\chi$ (and
therefore $N$) large, in order to enhance the strength of the localized term.
We thus have to relax the crystallographic restriction on the action
of the orbifold group, which will also break supersymmetry, but this is not
important for the purpose of our computation. Note that in this limit the
strength of the induced term in eq.~\echi\ $\chi_0\sim N$, since the sum over
$g$ and $h$ in eqs.~\eKKI\ and~\eFZ\ brings a factor of $N^2$.

To deduce the width of the localized term on the fixed-point, i.e. at $y=0$,
it is necessary to study the large-$q$ limit of the correction $\delta(q)$.
The problem is in finding the stationary points of $F_{(h,g)}$
in~\eFs. In the general $\ZZ_N$ orbifold case and in the large $N$ limit the
dominant contribution arises when $F$ approaches zero asymptotically. This can
be done by considering small values of $h\sim 1/N$ and sending $\tau_2$ to
infinity in an appropriate way.

In the large-$\tau_2$ limit, the oscillator contributions in the
expression~\eProp\ of $\chi_{ij}$ disappear and we are left with the zero mode
part:
\eqn\elnchi{
\ln\chi(z)\simeq -{\pi\over\tau_2}(\Im{\rm m}z)^2+\ln|2\sin\pi z|\, .
}
The positions $z_{1,2,3}$ of the graviton vertex operators
have to be chosen such that $F_{(h,g)}$ in~\eFs\ reaches a minimum.
It follows that $\Im{\rm m}(z_i-z_j)\sim\tau_2$ go to infinity and one is
left with the minimization of
\eqn\elnchias{
-\ln\chi(z_{12})+\ln\chi(z_{23})+\ln\chi(z_{31})\qquad ;\qquad
\ln\chi(z)\simeq -{\pi\over\tau_2}(\Im{\rm m}z)^2+\pi|\Im{\rm m}z|\, ,
}
where $z_{ij}=z_i-z_j$. Using the constraints $z_{12}+z_{23}+z_{31}=0$ and
$\Im{\rm m}z_i\in [0,\tau_2]$, one finds:
\eqn\zij{
\Im{\rm m}z_{12}=-{2\over 3}\tau_2\quad ,\quad \Im{\rm m}z_{23}=\Im{\rm
m}z_{31}={\tau_2\over 3}\, .
}
Note the similarity of these values with the Gross and Mende
configuration that extremizes the four-point amplitude, in which case the
positions of the vertices are separated by half the period of the
torus~\refs{\GrossAR}.

At the minimum~\zij\ and in the limit $\tau_2\to\infty$ and $h\to 0$, the
function $F_{(h,g)}$ becomes:
\eqn\Fas{\eqalign{
\left. F_{(h,g)}\right|_{\rm min} &\simeq {1\over h} \left[{4\over 9}\pi
h\tau_2 +1-{1\over 1 - \exp(2i\pi (g+h\tau))}-{1\over 1 - \exp(-2i\pi
(g+h\bar\tau))}\right]+{\cal O}(h)\cr
&={1\over h} \left[{4\over 9}\pi h\tau_2-{1\over 2}\coth(i\pi(g+h\bar\tau))
-{1\over 2}\coth(-i\pi(g+h\tau))
\right]+{\cal O}(h)
}}
up to terms exponentially suppressed in $\tau_2$. The terms of order $h$ come
from the expansion of the $\Psi$ functions in the vertex~\eDeltaN\ and are
independent of $\tau_2$. It is now clear that $h\tau$ can be chosen in a way
that the $1/h$ term vanishes, and thus $F_{(h,g)}$ becomes of order $h\sim
1/N$. Indeed, to leading order, $\pi h\tau_2$ is given as a solution of the
equation $4x/9=\coth x$. As a result, the width decreases as
\eqn\eWidth{
w \simeq l_s/N^{1\over2}\simeq l_P\, ,
}
and is given by the induced Planck length.

%%%%%%%%%%%%%%%%%%%%%%%%%%%%%%%%%%%%%%%%%%%%%%%%%%%%%%%%%%%%%%%%%%%%%%%%%%%%%%
\newsec{Discussion of the results}

In the previous section we argued that in the limit where the strength of the
localized gravity kinetic terms is much larger than the higher dimensional
gravitational scale, $M_P\gg M$ in eq.~\twoHE, their effective width is fixed by
their strength, i.e. the 4d Planck length, $w\sim l_P$. Although our
computation was done in a particular class of orbifold models where the tree
level correction was absent, the result confirms the field theory expectations
of previous works~\refs{\DG, \KTT} and in the following we will assume that it
remains valid in general.

We can now summarize our previous analysis for type II string theories on a 6d
non-compact internal space. The relevant gravitational kinetic terms are:
\eqn\twoHEf{
{M_s^8\over g_s^2}\int d^{10}x \sqrt{g}{\cal R}_{(10)} +
\chi M_s^2 f(g_s)\int d^4x \sqrt{g}{\cal R}_{(4)}\, ,
}
where for simplicity we omitted numerical factors and we introduced the
function $f(g_s)$ in order to treat simultaneously tree level and one-loop
contributions in (non-compact) Calabi--Yau manifolds and orbifolds. In the weak coupling limit
\eqn\fgs{
f(g_s)=-{c_0\over g_s^2}\pm c_1+\dots\, ,
}
where $c_0$ and $c_1$ are positive numerical constants given in~\eLocal, while
the dots stand for exponentially suppressed terms. In the case of $\ZZ_N$
orbifolds, $c_0$ vanishes and $\chi\sim N$ for large $N$. It follows that
$M_P^2=\chi f(g_s) M_s^2$, implying that $\chi f(g_s)\simeq 10^{32}$ for $M_s\simeq 1$
TeV. On the other hand, from eq.~\twoHE\ one has
\eqn\rcf{
r_c=\left[\chi g_s^2f(g_s)\right]^{1\over6}l_s\, ,
}
while from eq.~\Rc\ with $w\sim l_P$ one obtains the crossover scale
\eqn\Rcf{
R_c={r_c^3\over w^2}\sim g_s\sqrt{\chi f(g_s)}{l_s^3\over
l_P^2}=g_s{l_s^4\over l_P^3}
\simeq g_s\times 10^{32}\ {\rm cm}\, ,
}
for $M_s\simeq 1$ TeV. 
At 4d distances smaller than $R_c$, between points on the ``brane"
where gravity is localized, one recovers ordinary Newton's law, while at larger
distances gravity becomes ten-dimensional. Imposing $R_c$ to be larger than the
size of the universe, $R_c\simgt 10^{28}$ cm, one obtains that the string
coupling can be relatively small, $g_s\simgt 10^{-4}$, while the Euler number
must be huge (and negative if $c_0\neq0$): $|\chi|\sim g_s^2\times 10^{32}\simgt 10^{24}$.

Thus, the hierarchy is obtained mainly due to the large value of $\chi$, which
can be lowered only if one imposes a weaker bound on $R_c$, depending on our
actual knowledge of gravity at very large distance scales. Note that in the
case of $\ZZ_N$ orbifolds we studied in the previous section, $f(g_s)\sim{\cal
  O}(1)$ and $\chi\sim N$ determines completely the hierarchy:
$|\chi|\sim(l_s/l_P)^2\simeq 10^{32}$. It is worth noticing that adjusting
only one parameter $\chi$ and keeping $g_s$ of order unity, one can account
for the hierarchy and simultaneously obtain $R_c$ larger than the size of the
universe. Actually, having large Euler number implies only a large number of
closed string massless particles with no a-priori constraint on the observable
gauge and matter sector which has different origin in type II theories, as we
discuss below. As mentioned in section 2, in the case of Calabi--Yau manifolds,
$\chi$ counts the difference between the numbers of vector multiplets and
hypermultiplets, $\chi=\pm 4(n_V-n_H)$ for type IIA/B, and thus must be
negative/positive since $c_0$ of eq.~\fgs\ is not vanishing. On the other
hand, in the case of orbifolds, the contribution comes from $c_1$ requiring
always a surplus of closed string twisted vectors~\refs{\Kohlprath}. All these
particles are localized at the orbifold fixed points and should have
sufficiently suppressed gravitational-type couplings, so that their presence with such a huge multiplicity does not contradict observations.

Note that these results depend crucially on the scaling of the width $w$ in
terms of the Planck length: $w\sim l_P$, implying $R_c\sim 1/l_P^{2\nu +1}$
(in string units) for $\nu=1$. If there are models with $\nu>1$, the required
value of $\chi$ would be much lower, becoming ${\cal O}(1)$ for $\nu\ge 3/2$.
In this case, the hierarchy would be determined by tuning the string coupling
to infinitesimal values, $g_s\sim 10^{-16}$, in analogy to the compact string
models studied in refs.~\refs{\AP, \ADG}.
An alternative way to avoid introducing large $\chi$ is to compactify the six
internal dimensions at a length scale $l_c\simlt 10^{16}$ cm. Indeed, for
$\chi$ of order unity, eq.~\Rcf\ with $f\sim{\cal O}(1/g_s^2)$ leads a
crossover scale of a size of the solar system: $R_c\sim l_s^3/l_P^2\simeq
10^{16}$ cm. Then, at distances larger than $l_c$ KK modes decouple, while at
distances shorter than $l_c$ localized terms dominate. As a result, Newton's
law remains four-dimensional at all scales~\refs{\KTT}.

A similar situation is obtained in the context of M theory. As discussed in
section 2, localization of gravity now arises in five dimensions and, upon
compactification in four, the relevant kinetic terms become:
\eqn\twoHEfM{
M^9 R_{11}\int d^{10}x \sqrt{g}{\cal R}_{(10)} +
\chi M^3 R_{11}\int d^4x \sqrt{g}{\cal R}_{(4)}\, ,
}
where $M=l_M^{-1}$ is the M theory scale and omitted numerical factors can be
trivially restored from eq.~\eLocalM. It follows that 
$\chi R_{11}/l_M\sim(l_M/l_P)^2\simeq 10^{32}$ for $M\simeq 1$ TeV, while
$R_c\sim\sqrt\chi l_M^3/l_P^2$ implying again
$\chi\simgt 10^{24}$ and $R_{11}/l_M\simlt 10^8$, or equivalently a
compactification scale for the eleventh dimension bigger than about 10 keV.

Finally, we would like to discuss the question of matter localization.  In a
realistic model with six non-compact dimensions and four-dimensional
localized gravity, one should also have matter and gauge interactions
localized at the same point of the non-compact Calabi--Yau manifold. One
possibility is to use D3 branes located at this point. In this case, the
string coupling is fixed by the gauge coupling $g_s\sim g_{\rm YM}^2\sim{\cal
O}(1)$ and thus the hierarchy should be accounted by the Euler number. Another
possibility is to consider singular points where massless charged states can
arise as D-branes wrapped around the collapsing cycles. However, although
chiral matter indeed comes localized in particular singular points of the
internal space, gauge fields are generically localized only in co-dimension
four surfaces, and thus propagate in two extra dimensions. This can be
understood by considering for instance CY threefolds obtained as $K3$
fibrations on a two-dimensional base. Massless gauge fields then arise by
wrapping D-branes around $K3$ singularities and propagate freely along the
base. In this case, one is forced to consider only four non-compact dimensions
and compactify the remaining two at the string scale.

On the other hand, models with gauge fields localized at co-dimension six
singularities might emerge in more general CY spaces, as suggested by the
analysis of ref.~\refs{\CandelasNC}. Another way to avoid this problem in
models with small string coupling, is to identify the gauge sector with D4
branes stretched between two parallel (Neveu-Schwarz) NS5 branes located at
the same point where gravity is localized. The gauge coupling is then
independent of $g_s$ and is given, instead, by the coupling of the little
string theory, $g_{lst}=g_sl_s/L=g_{\rm YM}^2$, in the limit $g_s, L\to 0$,
with $L$ the separation of the two NS5 branes.  The resulting models are very
similar to those examined in ref.~\refs{\ADG} in the context of TeV little
strings.

In conclusion, we have studied gravity localization and non-compact flat
dimensions in string theory and proposed possible realizations in a consistent
perturbative framework. It would be interesting to study in detail the
phenomenological consequences of these models and compare to other
realizations of TeV strings with compact dimensions.

\vfill
%%%%%%%%%%%%%%%%%%%%%%%%%%%%%%%%%%%%%%%%%%%%%%%%%%%%%%%%%%%%%%%%%%%%%%%%%%%%%%
{\bf Acknowledgements:} We are grateful to Carlo Angelantonj for many useful
discussions and  patient explanations of the material in
\refs{\CarloReview}. Conversations and correspondence with Costas Bachas,
Karim Benakli, Savas Dimopoulos, Elias Kiritsis and Wolfgang Lerche are also
gratefully acknowledged.  I.A.  thanks particularly Gregory Gabadadze for
discussions and collaboration in the initial stage of this work. 
R.M. thanks the Theory Division of CERN,
and P.V. thanks the Laboratoire de Physique Th{\'e}orique de l'ENS de Lyon for
hospitality during the course of the work. This work was
supported in part by the European Commission under the RTN contract
HPRN-CT-2000-00122 and the INTAS contracts 55-1-590 and 00-0334.

%%%%%%%%%%%%%%%%%%%%%%%%%%%%%%%%%%%%%%%%%%%%%%%%%%%%%%%%%%%%%%%%%%%%%%%%%%%%%%
\vfill\supereject
\appendix{A}{Tadpole cancellation condition}

We rederive the tadpole cancellation condition for the type I models on
$T^2\times K3$ studied in ref.~\refs{\AntoniadisB} since it is intimately connected
with the calculations in Section 2.

\bigskip
The procedure follows the methods explained in the standard
texts~\refs{\CarloReview,\GimonRQ}. Note that the definition of the open
string proper-time $t$ and the closed string proper-time $\tau$ differ from
the conventions of the papers~\refs{\AntoniadisB,\GimonRQ,\BachasZT} where
similar calculations were performed.

The partition function of the model receives contributions from the torus
${\cal T}$, the Klein bottle ${\cal K}$, the annulus ${\cal A}$ and the M{\"o}bius
strip ${\cal M}$
$$
{\cal Z} = {\cal T} + {\cal K}+  {\cal A}+  {\cal M}\ .
$$
Since the fundamental domain of the torus has a natural ultraviolet cutoff, the
torus amplitude does not contribute to the tadpole cancellation condition.
In the other surfaces, the general form of the one-loop amplitude is
\refs{\CarloReview,\GimonRQ}:

\eqn\eZdef{
{\cal Z}_\sigma = C_\sigma \times     \int_0^\infty {dt\over t} \, {1\over t^2}
\, {1\over \eta^2(\tau)} \, {P^{(2)}(t)\over \eta^2(\tau)} \times 
\sum_{s=1,2,3,4} (-1)^{s-1} {\theta_s^2(\tau)\over \eta^2(\tau)} \times     Z_s(\tau)\, ,
}
where $C_\sigma$ are normalization constants depending on the Riemann
surface: $C_{\cal A}=1/4$, $C_{\cal M}=-1/4$ and $C_{\cal K}=1/2$.  $Z_s$ is
the $K3$ partition function including the zero-modes, while the sum is
performed over all the spin structures $s$. In the main
text and the following of this appendix, we follow the conventions
of~\refs{\AntoniadisB} where the internal partition function
contains the symmetry factor $C_\sigma$, so that 
${\rm Z}_{s,\sigma}(\tau) = C_\sigma \,Z_s(\tau)$.
 $\theta_s^2/\eta^2$ is the contribution of the
fermionic oscillators associated with the $T^2$,
$P^{(2)}/\eta^2$ represents the bosonic contribution from KK states and
oscillators of $T^2$, and the $1/t^2\eta^2$ comes from the non-compact
coordinates. All zero-mode parts are evaluated in terms of the proper-time
$t$, while oscillator contributions are computed on the double-cover of the
``open" Riemann surfaces and are functions of the closed string proper-time
$\tau$.
The $d$-dimensional KK-lattice sum $P^{(d)}$ and winding-lattice
sum $W^{(d)}$ are given by:
$$
P^{(d)}(t) = \sum_{p\in {}^*\Gamma_d} e^{- \pi t p^2}, \qquad W^{(d)}(t) =
\sum_{w\in \Gamma_d} e^{- \pi t w^2}\, ,
$$
with $p^2=|m+nU|^2/T_2U_2$ and $w^2=T_2|m+nU|^2/U_2$ for $d=2$.

\item{$\triangleright$} The Klein bottle vacuum amplitude reads:

$$
{\cal K} = \int_0^\infty {dt\over t^3} \, {P^{(2)}(t)\over
\eta^4(2it)} \,  \sum_{s=1,2,3,4}
(-1)^{s-1} {\theta_s^2(2it)\over \eta^2(2it)} \times    {\rm Z}_{s,{\cal K}}(2it)
$$
In order to extract the tadpole, we have to convert the amplitude to
the closed  string channel and express the divergence in terms of the closed
string proper-time $\ell$. The latter is given by  $2it = \tau \to -1/\tau 
= i\ell$ \eOtoC.  Now we can
extract the tadpole contribution:
$$
\tilde{\cal K} =  2^3\,T_2\,\int^\infty_0 d\ell {W^{(2)}(2l)\over
\eta^4(i\ell)} \times    \sum_{s=1,2,3,4}
(-1)^{s-1} {\theta_s^2(i\ell)\over \eta^2(i\ell)} \times    \tilde{\rm Z}_{s,{\cal K}}(i\ell) \ ,
$$
which together with the sum over the spin structure~\eSumSpin\ valid for $K3$
models gives 
\eqn\eKtadpole{
\left.\tilde{\cal K}\right|_{\rm tadpole} =   T_2\, (1-1) \,2^3\,
{\rm Z}_{s=1,{\cal K}}^{int}\, \int^\infty d\ell  \ .
}

\item{$\triangleright$}The annulus amplitude is  given by

$$
{\cal A} =  \int_0^\infty {dt\over t^3}\, {P^{(2)}(t)\over \eta^4(i
t/2)} \,  \sum_{s=1,2,3,4} (-1)^{s-1} {\theta_s^2(it/2)\over \eta^2(it/2)} \times 
{\rm Z}_{s,{\cal A}}(it/2) 
$$
After transforming the amplitude into the closed string channel using~\eOtoC: $
it/2 = \tau \to -1/\tau = i\ell$, and  summing over the spin
structure~\eSumSpin\ for $K3$ model, one can extract the tadpole
contribution:
\eqn\eTildeAnnulus{
\left.\tilde {\cal A}\right|_{\rm tadpole} = T_2\,(1-1)\, {1\over 2^3}\,  {\rm Z}_{s=1,{\cal A}}^{int}\,\int^\infty d\ell  \ .
}

\item{$\triangleright$} The M{\"o}bius strip amplitude is given by

$$
{\cal M} =  \int_0^\infty {dt\over t^3} \,   {P^{(2)}(t)\over \eta^4(1/2+i t/2)} \,  \sum_{s=1,2,3,4}
(-1)^{s-1} {\theta_s^2(1/2+it/2)\over \eta^2(1/2+it/2)} \times   {\rm Z}_{s,{\cal M}}(1/2+it/2) 
$$
Transforming again the amplitude into the closed string channel using~\eOtoC: 
$1/2+ it/2 =\tau \to -1/\tau \to -1/\tau+2 \to (1/\tau-2)^{-1}= -1/2+i/2t =
i\ell -1/2$, and using the sum over the spin
structure~\eSumSpin\ for $K3$ model, we extract the corresponding tadpole
contribution:
\eqn\eTadpoleM{
\tilde{\cal M}=  T_2\,(1-1)\, {1\over2}\,{\rm Z}_{s=1,{\cal M}}^{int}\,\int^\infty d\ell  \ .
}

\noindent
Tadpole cancellation implies the vanishing of the sum of the coefficient of
the integral:
\eqn\eTadpole{
{1\over 8}\, {\rm Z}_{s=1,{\cal A}}^{int}+2\,{\rm Z}_{s=1,{\cal
M}}^{int}+8\,   {\rm Z}_{s=1,{\cal K}}^{int}=0
}
which is the coefficient found in~\edeltaTwo. Note the absence of the $K3$
volume in the above tadpole condition, following from the fact that we
considered only the odd spin structure $s=1$ associated with a localized term.
In the orbifold limit, it reduces to the twisted tadpole which has to vanish
locally at the fixed points even in the non-compact case.

\vfill\supereject
%%%%%%%%%%%%%%%%%%%%%%%%%%%%%%%%%%%%%%%%%%%%%%%%%%%%%%%%%%%%%%%%%%%%%%%%%%%%%%
\appendix{B}{Untwisted vertex operator emission for orbifolds}

In this appendix we describe some properties on the construction of 
the untwisted closed string vertex operator $V_{(p,\omega)}$~\eVTTU\
used in the main text. For simplicity, we restrict ourselves to one
orbifold coordinate $S^1/\ZZ_N$.

The various sectors under the orbifold action will be labelled by $h$ and
$g$. $h$ corresponds to twisted states running around the loop, {\it i.e.\/}
taking a trace over the twisted states $t$, $\sum_t \langle t| \cdots
|t\rangle$ and $g$ to the insertion of the orbifold projector inside the
trace, {\it i.e.\/} ${\rm Tr}( \gamma^g \cdots )$.

\subsec{Definitions and notations}

\noindent
This vertex operator is constructed in a similar way as for the heterotic
string~\refs{\GreenMN}, using the $\ZZ_N$ twisted coordinate $X$ under the
orbifold action~\refs{\HamidiVH}:
$$
V_{(p,\omega)}= :e^{i p_L X_L + i p_R X_R}:\ ,
$$
with the coordinates twisted by the orbifold action $\exp(2i\pi\, l\, v/N)$. $X_{L,R}$ have the following mode expansion
\refs{\GinspargUI} with $h=l \,v/N$ 
%\eqn\eXLR{
$$\eqalign{
X_L &= x_L + i \sqrt{2\alpha^\prime}\sum_{n\in\ZZ} \left( {\alpha_{n-h} \over n-h} z^{-(n-h)} +
{\alpha_{n+h}^\dagger\over n+h}  z^{-(n+h)}\right)\cr
X_R &= x_R + i\sqrt{2\alpha^\prime} \sum_{n\in\ZZ} \left( {\tilde \alpha_{n+h} \over
n-h} \bar z^{-(n-h)} +{\tilde \alpha_{n-h}^\dagger\over n+h}  \bar z^{-(n+h)}\right)\ .\cr
}
$$
Here, $x_f=(x_L+x_R)/\sqrt 2$ is the fixed point $x_f=f\, 2\pi R/(1-\gamma)$,
with $f=0,1$ and $\gamma=-1$ for a $\ZZ_2$-orbifold. The quantization of the
oscillators implies:
$$
[\alpha_{n-h}, \alpha_{m+h}^\dagger]= 
[\tilde \alpha_{n-h}, \tilde\alpha_{m+h}^\dagger] =(m+h) \delta_{n+m,0}\, ,
$$
while the zero modes satisfy~\refs{\ErlerAN}:
$$
[x_L , x_R] = i {2\pi\over 1-\gamma}\ .
$$
The standard definition of the normal ordering gives the following split
between the zero mode and oscillator parts:
\eqn\eVsplit{
V_{(p,w)}= \Delta_h^{-\alpha^\prime(p^2+\omega^2)} e^{i (p_L {X}_L^{(-)} + p_R
X_R^{(-)})} e^{i (p_L x_L + p_R x_R)} e^{i (p_LX_L^{(+)} + p_R X_R^{(+)})}\, ,
} 
with $X_{L,R}^{(-)}$ and $X_{L,R}^{(+)}$ the negative and positive frequency
parts of $X_{L,R}$, respectively.

\indent{$\triangleright$} The zero-mode part of~\eVsplit\ is given by $\exp(i (p_L x_L + p_R x_R))$. 
Changing variables to $(x_f,q)$ with $x_{L/R} = (x_f\pm       q)/\sqrt 2$, we
get the canonical commutation relation

$$
[x_f,q] = i{2\pi\over 1-\gamma} \ ,
$$
and the zero-mode part takes the form \refs{\ErlerAN}:
$$
e^{i (p_L x_L + p_R x_R)} = e^{-i\pi p\omega}\  e^{ip x_f }\ e^{i\omega q}\ .
$$
Since $q$ is the canonical conjugate variable to the position of the fixed
point, the last term is just the shift operation of the position of the fixed
point (this is the operator $S(\omega)$ introduced in the definition of the
vertex operator in~\refs{\HamidiVH}):
$$
e^{i \omega q} |x_f\rangle = |x_{f+\omega}\rangle\ .
$$

\indent{$\triangleright$}  The standard definition of normal
ordering \refs{\GreenMN} does not include the factor $\exp(-\alpha^\prime
p^2\sum_n 1/n)$ from the commutator $\exp([X^{(-)},X^{(+)}]/2)$. According to
this definition we find a normal ordering prefactor, on the cylinder, given by:

$$\eqalign{
e^{{1\over 2} \left(p_L^2 [{X}_L^{(-)},{X}_L^{(+)}] + p_R^2
[{X}_R^{(-)},{X}_R^{(+)}] \right)}\times     e^{\alpha^\prime
(p_L^2+p_R^2) \sum_n^\prime {1\over n}} &= e^{- \alpha^\prime(p_L^2+p_R^2)
\sum_n {1\over |n-h|} - \sum_n^\prime {1\over |n|}}\cr & =
\Delta_h^{-\alpha^\prime (p^2+\omega^2)}\ . }
$$
The factor $\Delta_h$ defined above is the ratio of the normal ordering
factor between twisted and untwisted states~\refs{\HamidiVH}.  It can be
obtained as the difference between the coincident point correlators $X^2(0)$
in the twisted and untwisted sector:
\eqn\eDeltaDef{
\ln\Delta_h= \left\langle X(0) X(0)\right\rangle_{\rm twisted} - 
\left\langle X(0) X(0)\right\rangle_{\rm untwisted}\, .
}
This definition will allow us to derive the corresponding $\ln\Delta_h$
factor on the torus.

We have to compute the vacuum expectation value $\langle
V_{(p,\omega)}(z_3)\rangle$ of the vertex operator~\eVsplit, associated with
the emission of an untwisted state from the twisted sectors. A naive
evaluation of this operator using Wick theorem would set this expectation value
to zero, but the correct answer is
\eqn\eVev{
\left.\left\langle V_{(p,\omega)}(z_3)\right\rangle\right|_{h}
=\Delta_h^{-\alpha^\prime(p^2+\omega^2)} \, \sum_{\rm  fixed\ points} 
e^{i p x_f}\, .
}
In the following, we derive this correlator from a massive deformation of the
partition function of a free (twisted) boson.

%%%%%%%%%%%%%%%%%%%%%%%%%%%%%%%%%%%%%%%%%%%%%%%%%%%%%%%%%%%%%%%%%%%%%%%%%%%%%
\subsec{The twisted correlator on the torus}

\noindent
On the torus, the twisted coordinates are expanded as
$$
X = x_f + \sum_{m,n\in\ZZ} \, \alpha_{n+h,m+g}\,
\phi_{n+h,m+g}(\tau,z)\, ,
$$
with $\phi_{n+h,m+g}(\tau,z)$ a basis of eigen-functions of $-\Delta_z$ with
the eigenvalue $\lambda_{n+h,m+g}=(2\pi)^2|(n+h)\tau + m+g|^2/\tau_2^2$ for the
boundary conditions specified by $h$ and $g$~\refs{\doph}. The coincident
point correlator is extracted from the
partition of a twisted boson with mass $(2\pi\mu )^2$ after differentiating with
respect to $(2\pi\mu )^2$. 
The eigenvalues of the massive Laplace operator $-\Delta_z +(2\pi\mu )^2$ are
$\lambda_{n+h,m+g}=(2\pi)^2\left[|(n+h)\tau + m+g|^2/\tau_2^2+\mu ^2\right]$,
and the partition function, computed using the standard zeta-function
regularization following the references~\refs{\doph,\Itzykson}, is
\eqn\Zmassive{\eqalign{
{\cal Z}(h,g;\mu ^2)&=\int {\cal D}X e^{-\int_{{\cal T}^2} \, d^2z\, X (-\Delta_z +(2\pi\mu )^2) X} =
\prod_{n,m} {1\over\lambda_{n+h,m+g}}\cr
&=e^{2\pi \tau_2 \gamma_h} \, \left|\prod_{n\in\ZZ}
\left(1-e^{2i\pi ( g + (n+h)\tau_1)-2\pi \tau_2 \sqrt{(n+h)^2 +\mu ^2}}\right)\right|^{-2}\, .
}}
The prefactor depends on the mass $\mu ^2$ and the twist $h$ only, and
represents the zero-point energy of the massive boson~\refs{\Green,\kata}
$$
\gamma_h = {1\over 2\pi} \sum_{n\neq0} \int_0^\infty dt \,
e^{-\pi t n^2-\pi \mu ^2/t + 2i\pi n h}\, .
$$
We will need its expansion with respect to $\mu ^2$, given in~\refs{\Itzykson}:
$$
\eqalign{
\gamma_{h\neq0} & =\left(h-{1\over2}\right)^2- {1\over 12}+ {\mu ^2\over2} (\ln(4\pi)^2 + \Psi(h)+\Psi(1-h))+ {\mu ^4\over4}
\int_0^1 dl  \sum_{n\in\ZZ} {(1-l)\over ((n+h)^2 +l \mu ^2)^{3\over2}}\cr
\gamma_0 & = {1\over 6} - \mu + {\mu ^2\over2} (\ln(4\pi)^2 + 2\Psi(1))+ {\mu ^4\over4}
\int_0^1 dl  \sum_{n\neq0} {(1-l)\over (n^2 +l \mu ^2)^{3\over2}}\, .}
$$
The correlator $\langle X(0) X(0)\rangle$ in the $(h,g)$ sector is deduced by
differentiating the logarithm of the massive partition
function~\Zmassive, and $\ln\Delta_h$ is given by
\eqn\eDeltahNew{
\ln\Delta_h =\lim_{\mu ^2\to0}\left( -{1\over\pi\tau_2}\, \partial_{\mu ^2} \ln
{\cal Z}(h,g;\mu ^2) +\ln(4\pi)^2 + 2\Psi(1)\right)  \, .
}
\vfill\break
\indent$\triangleright$ The result for the twisted sector $h\neq0$ is

\eqn\eDeltahNZ{
\ln\Delta_{h\neq0} = 2\Psi(1)-\Psi(h)-\Psi(1-h)+ \sum_{n\in\ZZ\atop m>0}
{1\over|n+h|}\, e^{2i\pi m(g+(n+h)\tau_1) -2\pi m \tau_2 |n+h|}+c.c.
}
In the field theory limit $\tau_2\to\infty$, only the first
term survives, representing the tree-level coupling between twisted and
untwisted states obtained in~\refs{\HamidiVH,\DixonQV}. 

\indent{$\triangleright$} In the untwisted sector $h=0$ (but $g\neq0$)
we have

\eqn\eDeltahZero{
\ln\Delta_{h=0} = {\pi\tau_2\over2\sin(\pi g)^2}+\sum_{n>0\atop m>0}
{1\over n}\, \left( e^{2i\pi m(g+n\tau)}+e^{2i\pi m(g-n\bar\tau)}+c.c.\right) \, .
}
Because of the term linear in $\tau_2$, the amplitude vanishes in the
field theory limit $\tau_2\to\infty$, since the contributions from the
untwisted states $h=0$ running in the loop vanish due to momentum conservation.
For $g=0$, the expression is infinite due to the presence of a zero mode.
Actually, it is easy to check that the expression~\eDeltahZero\ is the $h\to0$
limit of~\eDeltahNZ.

\indent{$\triangleright$} In order to study the modular invariance of
these expressions we follow~\refs{\Green,\kata}, where
the modular properties  of 
${\cal Z}(h,g;\mu ^2|\tau)={\cal Z}(-g,h;\mu ^2|\tau||-1/\tau)$ where studied.
We consider the following integral representation:

\eqn\elnI{\eqalign{
\ln{\cal Z}(h,g;\mu ^2) &= 2\pi\tau_2\gamma_h + \sum_{n\in\ZZ\atop m>0}
{1\over m}\,
e^{2i\pi m(g+(n+h)\tau_1) -2\pi m \tau_2 \sqrt{(n+h)^2+\mu ^2}}+c.c.\cr
&= 2\pi \tau_2 \gamma_h + \sum_{n\in\ZZ\atop m>0}
 \int_0^\infty {dt\over t^{1\over2}} \, e^{-\pi t m^2 -{\pi \tau_2^2\over t} (
(n+h)^2+\mu ^2) +2i\pi m(g+(n+h)\tau_1)} + c.c.
}}
and perform a Poisson resummation on $n$:
\eqn\elnII{
\ln{\cal Z}(h,g;\mu ^2) = 2\pi \tau_2 \gamma_h +{1\over \tau_2} \sum_{\hat n\in\ZZ\atop m>0}
 \int_0^\infty dt \, e^{-\pi\tau_2^2 \mu ^2/t-\pi t m^2 -t\pi
(\hat n+m\tau_1)^2/\tau_2^2 +2i\pi (mg-\hat nh)} + c.c.
}
Differentiating with respect to $\mu ^2$, one finds:
$$
\eqalign{
{1\over 2\pi\tau_2}\partial_{\mu ^2}\ln{\cal Z}(h,g;\mu ^2)& = \partial_{\mu ^2}\gamma_h - {1\over2}\sum_{\hat n\in\ZZ\atop m>0}
 \int_0^\infty {dt\over t} \, e^{-{\pi\over t}\tau_2^2 \mu ^2-\pi t m^2 -t\pi
(\hat n+m\tau_1)^2/\tau_2^2 +2i\pi (mg-\hat nh)} + c.c.\cr
&= \partial_{\mu ^2}\gamma_h -{1\over2} \sum_{\hat n\in\ZZ\atop m\neq 0}
 \int_0^\infty {dt\over t} \, e^{-{\pi\over t}\tau_2^2 \mu ^2-\pi t m^2
-{t\pi\over \tau_2^2}(\hat n+m\tau_1)^2 +2i\pi (mg-\hat nh)}\cr
&= \partial_{\mu ^2}\gamma_h -{1\over2} \sum_{\hat n\in\ZZ\atop m\neq 0}
 \int_0^\infty {dt\over t} \, e^{-{\pi\over t}\tau_2^2 \mu ^2-{\pi
t|\tau|^2\over \tau_2^2} \left(m+\hat n {\tau_1\over|\tau|^2}\right)^2
-{t\pi\over \tau_2^2}{\hat n}^2 +2i\pi (mg-\hat nh)}\cr
&\mathop{\longrightarrow}_{\mu ^2\to 0} \partial_{\tilde \mu ^2}\gamma_g -\ln|\tau|\cr 
&-{\tau_2\over2|\tau|^2}\lim_{\mu ^2\to0} \sum_{\hat m\in\ZZ\atop \hat n\neq 0} \int_0^\infty
{dt\over t^{3\over 2}} \,
e^{ -\pi t {\hat n}^2-{\pi\over t} {\tau_2^2\over|\tau|^4}\left((\hat
m+g)^2 + \tilde\mu ^2\right)  -2i\pi \hat n\left(h+(g+\hat m) {\tau_1\over|\tau|^2}\right)}\cr
}
$$
The last equality has been obtained after Poisson resummation on
$m$ and we introduced $\tilde\mu ^2= |\tau|^2\mu ^2$. In the first line of the last equality, the modular ``anomaly'' arises
from the sector $\hat n=\hat m=0$~\refs{\Itzykson}. In the second line, by rescaling $t$ as $t\to |\tau|^2\,t$ and relabeling the
integers as $(\hat n,\hat m)\to (m,n)$, we bring this expression into the same
form as the $\tilde \mu ^2$-derivative of the second line of~\elnI\ with the
exchange of $h$ and $g$ together with $\tau \to -1/\tau$. And we conclude that
\eqn\eDeltMod{
\ln\Delta_h(-1/\tau) = \ln\Delta_g(\tau) - \ln|\tau|^2
}

\vfill
%%%%%%%%%%%%%%%%%%%%%%%%%%%%%%%%%%%%%%%%%%%%%%%%%%%%%%%%%%%%%%%%%%%%%%%%%%%%%%
\listrefs

\bye